\documentclass[iop, apj]{emulateapj}
\usepackage[english]{babel}
\usepackage[utf8]{inputenc}
\usepackage{graphicx}
\usepackage{amsmath}
\usepackage{mathtools}
\usepackage{natbib}
\usepackage{wasysym}

\defcitealias{2014MNRAS.441.1783A}{AM14}

\begin{document}

\title{The ALHAMBRA survey: evolution of galaxy spectral segregation}

\author{
Ll.~Hurtado-Gil\altaffilmark{1,2},
P.~Arnalte-Mur\altaffilmark{1},
V.~J.~Mart\'inez\altaffilmark{1,3,4},
A.~Fern\'andez-Soto\altaffilmark{2,4},
M.~Stefanon\altaffilmark{5},
B.~Ascaso\altaffilmark{6,21},  
C.~L\'opez-Sanju\'an\altaffilmark{8}, 
I.~M\'arquez\altaffilmark{7},
M.~Povi\'c\altaffilmark{7},
K.~Viironen\altaffilmark{8},
J.~A.~L.~Aguerri\altaffilmark{9,10},
E.~Alfaro\altaffilmark{7},
T.~Aparicio-Villegas\altaffilmark{11,7}
N.~Ben\'itez\altaffilmark{7},           
T.~Broadhurst\altaffilmark{12,13}, 
J.~Cabrera-Ca\~no\altaffilmark{14},
F.~J.~Castander\altaffilmark{15},
J.~Cepa\altaffilmark{9,10},
M.~Cervi\~no\altaffilmark{7,9,10},         
D.~Crist\'obal-Hornillos\altaffilmark{8}, 
R.~M.~Gonz\'alez~Delgado\altaffilmark{7},
C.~Husillos\altaffilmark{7},
L.~Infante\altaffilmark{16},          
J.~Masegosa\altaffilmark{7},  
M.~Moles\altaffilmark{8,7},
A.~Molino\altaffilmark{17,7},
A.~del~Olmo\altaffilmark{7},
S.~Paredes\altaffilmark{18},
J.~Perea\altaffilmark{7},              
F.~Prada\altaffilmark{7,19,20},
\and J.~M.~Quintana\altaffilmark{7}
}

\begin{abstract}
{\small
We study the clustering of galaxies as a function of spectral type and redshift in the range $0.35 < z < 1.1$ using data from the Advanced Large Homogeneous Area Medium Band Redshift Astronomical (ALHAMBRA) survey. The data cover 2.381  deg$^2$ in 7 fields, after applying a detailed angular selection mask, with accurate photometric redshifts [$\sigma_z < 0.014(1+z)$] down to $I_{AB} < 24$. From this catalog we draw five fixed number density, redshift-limited bins. We estimate the clustering evolution for two different spectral populations selected using the ALHAMBRA-based photometric templates: quiescent and star-forming galaxies. For each sample, we measure the real-space clustering using the projected correlation function. Our calculations are performed over the range $[0.03,10.0] h^{-1}$ Mpc, allowing us to find a steeper trend for $r_p \lesssim 0.2 h^{-1}$ Mpc, which is especially clear for star-forming galaxies. Our analysis also shows a clear early differentiation in the clustering properties of both populations: star-forming galaxies show weaker clustering with evolution in the correlation length over the analysed redshift range, while quiescent galaxies show stronger clustering already at high redshifts, and no appreciable evolution. We also perform the bias calculation where similar segregation is found, but now it is among the quiescent galaxies where a growing evolution with redshift is clearer.% (abrigatted).
These findings clearly corroborate the well known colour-density relation, confirming that quiescent galaxies are mainly located in dark matter halos that are more massive than those typically populated by star-forming galaxies.}
\end{abstract}

\section{Introduction}
\label{sec:intro} 

It has been well established that different types of galaxies cluster in  different ways \citep{1988ApJ...331L..59H,1988ApJ...333L...9D, 1989ApJ...339L...9D, 1991MNRAS.252..261E,1995ApJ...442..457L,1997ApJ...489...37G,2006MNRAS.368...21L,2010A&A...513A..22M, 2006A&A...457..145P}. 
Elliptical galaxies are preferentially located at the cores of rich galaxy clusters, i.e, in high density environments, while spiral galaxies are the dominant population in the field \citep{1976ApJ...208...13D,1980ApJ...236..351D, 1986ApJ...300...77G, 2006A&A...458...39C}.
This phenomenon, called galaxy segregation, has been confirmed in the largest galaxy redshift surveys available up to date, the 2dF Galaxy redshift survey \citep[2dFGRS,][]{2003MNRAS.344..847M}, the Sloan Digital Sky Survey \citep[SDSS,][]{2006MNRAS.372.1749A,2011ApJ...736...59Z} and the Baryonic Oscillation Spectroscopic Survey \citep[BOSS,][]{2013ApJ...767..122G}. The dependence of clustering on different galaxy properties such as stellar mass, concentration index, or the strength of the 4000 \AA -break has been studied by \cite{2006MNRAS.368...21L}.  

Since segregation is a consequence of the process of structure formation in the universe, it is therefore very important to understand its evolution with redshift or cosmic time. 
Several works have extended the analysis of segregation by colour or spectral type to redshifts  in the range $z \sim 0.3 - 1.2$ using recent spectroscopic surveys such as the VIMOS-VLT Deep Survey \citep[VVDS,][]{Meneux2006a}, the Deep Extragalactic Evolutionary Probe 2 survey \citep[DEEP2,][]{coi08a}, or the PRIsm Multi-object Survey \citep[PRIMUS,][]{Skibba2013c}. \Citet{tor09a}, instead, used the zCOSMOS survey to study segregation by morphological type at $z \sim 0.8$.
All these studies show that segregation by colour or spectral type was already present at $z \sim 1$. 
In particular,  \citet{Meneux2006a}, using a sample of 6,500 VVDS galaxies covering half a square degree, have unambiguously established that early-type galaxies are more strongly clustered than late-type galaxies at least since redshift $z\sim 1.2$. The correlation length obtained by these authors for late-type galaxies is $r_0 \sim 2.5 h^{-1}$ Mpc at $z \sim 0.8$ and roughly twice this value for early-type galaxies. They have also calculated the relative bias between the two types of galaxies obtaining an approximately constant value $b_{\rm rel} \sim 1.3-1.6$ for $0.2 \le z \le 1.2$ depending on the sample. This value is slightly larger than the one obtained by \citet{2003MNRAS.344..847M} $b_{\rm rel} \sim 1.45 \pm 0.14$ for the 2dF Galaxy Redshift Survey with median redshift $z=0.1$. The results obtained by \citet{coi08a} for DEEP2 reinforced those outlined above, although the measured correlation lengths for DEEP2 galaxies are systematically slightly larger than the values reported for the VVDS sample by \citet{Meneux2006a}. In addition, \citet{coi08a} have detected a significant rise of the correlation function at small scales $r_p \le 0.2 h^{-1}$ Mpc for their brighter samples.
For the zCOSMOS-Bright redshift survey, \citet{tor09a} found also that early-type galaxies exhibit stronger clustering than late-type galaxies on scales from $0.1$ to 
$10 h^{-1}$ Mpc already at $z \simeq 0.8$, and the relative difference increases with cosmic time on small scales, but does not significantly evolve from $z=0.8$ to $z=0$ on large scales. A similar result is reported by \citet{Skibba2013c}. These authors show that the clustering amplitude for the PRIMUS sample increases with color, with redder galaxies displaying stronger clustering at scales 
$r_p \leq 1 h^{-1}$ Mpc. They have also detected a color dependence within the red sequence, with the reddest galaxies being more strongly correlated than their less red counterparts. This effect is absent in the blue cloud.

Several broad-band photometric surveys have extended these studies to even larger redshift \citep[e.g.][]{har10a, McCracken2015c}. \citeauthor{har10a}, using data from the UKIDSS Ultra Deep Survey, find segregation between passive and star-forming galaxies at $z \lesssim 1.5$, but find consistent clustering properties for both galaxy types at $z \sim 2$.

In the present paper we use the high-quality data of the Advanced Large Homogeneous Area Medium-Band Redshift Astronomical survey (ALHAMBRA) \citep{mol08a, 2013arXiv1306.4968M}\footnote{http://alhambrasurvey.com} to study the clustering segregation of quiescent and star-forming galaxies. ALHAMBRA is very well suited for the analysis of galaxy clustering and segregation studies at very small scales. With a reliable calculation of the projected correlation function we find a clear steepening of the correlation at scales between $0.03$ to $0.2 h^{-1}$ Mpc \citep{2006A&A...457..145P, coi08a}, specially for the star-forming galaxies. Moreover, its continuous selection function over a large redshift range makes ALHAMBRA an ideal survey for evolution studies. In \cite{2014MNRAS.441.1783A} (hereafter \citetalias{2014MNRAS.441.1783A}) the authors presented the results of the evolution of galaxy clustering on scales $r_p < 10 h^{-1}$ Mpc for samples selected in luminosity and redshift over $\sim 5$ Gyr by means of the projected correlation function $w_p(r_p)$. 
In this paper we use the same statistic to study the evolution of galaxy segregation by spectral type at $0.35 < z < 1.1$.

Details on the samples used in this analysis are described in Section~\ref{sec:seg}.
In Section~\ref{sec:methods}, we introduce the statistic used in our analysis, the projected correlation function, and the methods to obtain reliable estimates of this quantity and to model the results. Finally, in Section~\ref{sec:res}, we present our results, and in Section~\ref{sec:con} the conclusions. 
Throughout the paper we use a fiducial flat $\Lambda CDM$ cosmological model with parameters $\Omega_M = 0.27$, $\Omega_{\Lambda} = 0.73$, $\Omega_b = 0.0458$ and $\sigma_8 = 0.816$ based on the 7-year \textit{Wilkinson Microwave Anisotropy Probe (WMAP)} results \citep{2011ApJS..192...18K}. 
All the distances used are comoving, and are expressed in terms of the Hubble parameter $h \equiv H_0/100 \rm{\ km} \rm{\ s}^{-1} \rm{\ Mpc}^{-1}$. Absolute magnitudes are given as $M - 5 \log_{10}(h)$.

\section{ALHAMBRA galaxy samples}
\label{sec:seg} 

The Advanced Large Homogeneous Area Medium-Band Redshift Astronomical
survey (ALHAMBRA) \citep{mol08a, 2013arXiv1306.4968M} is a project
that has imaged seven different areas in the sky through a
purposedly-built set of 20 contiguous, non-overlapping, 310 \AA-wide
filters covering the whole visible range from 3500 to 9700 \AA, plus
the standard near-infrared $JHK_s$ filters. The nominal depth (5$\sigma$, $3''$ aperture) is $I_{AB} \sim 24.5$ and the total sky coverage after masking is 2.381
deg$^2$. The final catalogue, described in \cite{2013arXiv1306.4968M}, includes over 400,000 galaxies, with a photometric redshift accuracy better than $\sigma_z / (1+z) =
0.014$. Full details on how the accuracy depends on the sample magnitude, 
galaxy type, and bayesian odds selection limits are given in that work. For the characteristics 
of the sample that we use in this paper the authors quote a dispersion $\sigma_{\rm NMAD} <
0.014$ and a catastrophic rate $\eta_1 = 0.04$\% \footnote{Where $\sigma_{\rm NMAD}$ is the normalized median absolute deviation, and $\eta_2$ is defined as the proportion of objects with absolute 
deviation $|\delta z| / (1+z) > 0.2$.}. There is no evidence of 
significantly different behaviour for galaxies with spectral energy distributions 
corresponding to quiescent or star-forming types. Contamination by AGNs is minor 
(approximately 0.1\% of the sources could correspond to this class, which has not been 
purged from the ALHAMBRA catalogues) and should be 
dominated by low-luminosity AGN, which are in many cases fit by strong emission-line galaxies 
with an approximately correct redshift.

Object detection is performed over a synthetic image, created
via a combination of ALHAMBRA filters, that mimics the Hubble Space
Telescope F814W filter (hereafter denoted by $I$) so that the
reference magnitude is directly comparable to other
surveys. Photometric redshifts were obtained using the
template-fitting code BPZ \citep{2000ApJ...536..571B}, with an
updated set of 11 Spectral Energy Distribution (SED) templates, as
described in \cite{2013arXiv1306.4968M}. Although a full posterior
probability distribution function in redshift $z$ and spectral type
$T$ is produced for each object, in this work we take a simpler
approach and assign to each galaxy the redshift $z_b$ and type $T_b$
corresponding to the best fit to its observed photometry. We have checked that the errors induced by the redshift uncertainties, which
are partly absorbed by the deprojection technique, are under control
as long as we use relatively bright galaxies with good quality
photometric redshift determinations. This makes ALHAMBRA a very well suited catalogue: together with the high resolution photometric redshifts, the abundant imaging allows us both a reliable color segregation, used in this work, and a high completeness in the galaxy population at small scale separations, which will be the specific object of a future work. 

We have drawn different samples from the ALHAMBRA survey to perform our analysis in a similar way as was done in \citetalias{2014MNRAS.441.1783A}. 
First, we cut the magnitude range at $I < 24$, where the catalogue is photometrically complete \citep{2013arXiv1306.4968M} and we do not expect any significant field-to-field variation in depth. 
Second, stars are eliminated using the star-galaxy separation method described in \cite{2013arXiv1306.4968M}.
As explained in \citetalias{2014MNRAS.441.1783A}, the expected contamination by stars in the resulting samples is less than 1~per cent.
Finally, we cleanse the catalogue using the angular masks defined in \citetalias{2014MNRAS.441.1783A}, which eliminate regions with less reliable photometry around bright stars or image defects, or very close to the image borders.
The sample selected in this way contains 174633 galaxies over an area of $2.381 \deg^2$, {\it i.e.}, with an approximate source density of $7.3 \times 10^4$ galaxies per square degree.

\begin{figure}
\begin{center}
\includegraphics[height=0.25\textheight]{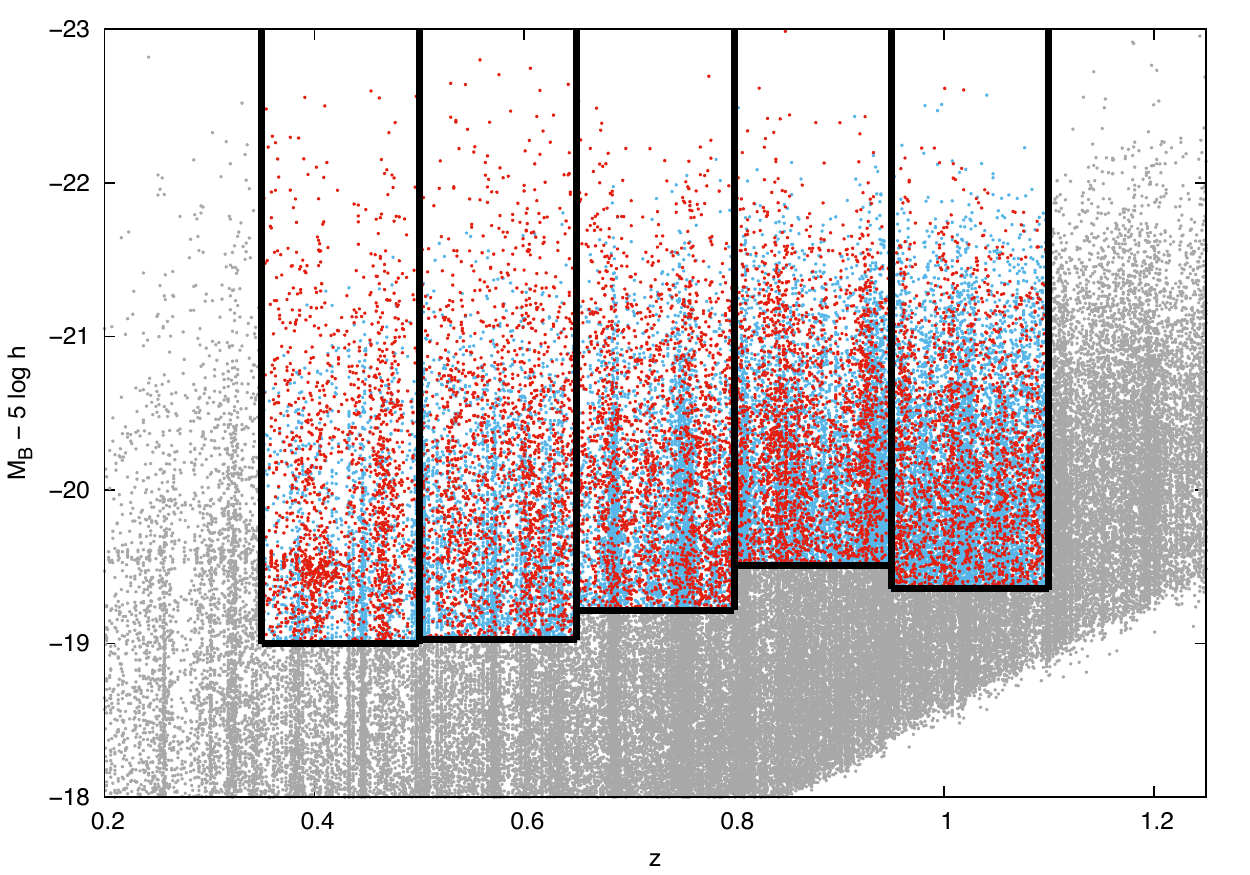}\\ 
\end{center}
\caption{\label{redmag}Selected samples with fixed number density in the photometric redshift vs. absolute $B$-band magnitude diagram. The quiescent and star-forming galaxy samples are plotted in red and blue color, respectively. The solid lines mark the boundaries of our selected samples described in Table~\ref{population}.}
\end{figure}

Given the ALHAMBRA depth, we divide our sample in 5 non-overlapping redshift bins.  These redshift bins are $[0.35,0.5[$, $[0.5,0.65[$, $[0.65,0.8[$, $[0.8,0.95[$, and $[0.95,1.1]$ \footnote{Note that the redshift bins used here are different to those in \citetalias{2014MNRAS.441.1783A}, where overlapping bins were allowed.}. 
As in this work we focus on the galaxy spatial segregation by spectral type we use a luminosity selection to obtain a fixed number density. In this way we guarantee that we are comparing similar populations at different redshifts. 
In order to select a sample that is complete up to $z = 1.1$ we define a threshold magnitude of $M_B^{th}(0) - 5\log(h) = -19.36$ for the highest redshift bin. This limit determines the galaxy number density ($\bar{n} = 9.35 \times 10^{-3} \, h^{3}$ Mpc$^{-3})$ that we will keep constant for the remaining redshift bins. This way, our results do not rely on measurements of the luminosity function. This will allow us to study the evolution with redshift of the galaxy spectral segregation. Figure ~\ref{redmag} shows the luminosity and redshift selections used in this work. We should remark on the non-monotonic evolution of the faint limit of our samples with redshift. This effect is not unexpected, as a combination of cosmic variance in the large-scale structure and the artificial redshift peaks that are induced by the photometric redshift methods produce density changes that are observable at the scales we are using. In any case the effect is very small, representing a variation of only 0.1 magnitudes per bin over a monotonic evolution.

We classify our galaxies as `quiescent' and `star-forming' according to the best-fitting template, $Tb$, obtained from the BPZ analysis.
Templates 1 to 5 correspond to quiescent galaxies, 6 and 7 correspond to star-forming galaxies, and 8 to 11 correspond to starburst galaxies. We consider as quiescent galaxies those with a template value smaller than 5.5, and star-forming those with a value bigger than 5.5. Therefore, we include in the star-forming category also those galaxies classified as starbursts. Note that in the fitting process interpolation between templates is performed. 

In a previous work \cite{2013MNRAS.435.3444P}  built a morphological catalogue of 22,051 galaxies in ALHAMBRA. We cannot, however, use this catalogue as the basis for our analysis as it includes only a small subset of the galaxies in our sample: in its cleanest version it is limited to $AB(\rm{F613W})<22$ and redshift $z < 0.5$ for ellipticals. A cross-check showed that, if we identify quiescent galaxies as early-type and star-forming galaxies as late-type, our SED-based classification agrees with the morphological one for over 65\% of the sample. Taking into account that the nominal accuracy of the morphological catalogue is 90\%, that we are actually using only the objects close to its detection limit and that, as noted in \cite{2013MNRAS.435.3444P}, the relationship between morphological- and colour-based classifications is far from being as direct as could na\"ively be expected, we consider that these figures prove that the classification is accurate within the expected limits.

\begin{deluxetable*}{lcc|cccc|cccc|c}
\tablewidth{0pt}
\tablecaption{\label{population}
Characteristics of the galaxy samples used 
}
\tablehead{
\colhead{}       & \colhead{}          & \colhead{}                   & \multicolumn{4}{c}{Quiescent galaxies}                                                                      & \multicolumn{4}{c}{Star-forming galaxies} & \colhead{}\\
\colhead{Sample} & \colhead{$z$ range} & \colhead{$V (h^{-3} Mpc^3)$}  & \colhead{$N_{\rm Q}$} & \colhead{$\bar{n} (h^{3} Mpc^{-3})$} & \colhead{$M_{B}^{\rm med}$} & \colhead{$\bar{z}$}    & \colhead{$N_{\rm Sf}$} & \colhead{$\bar{n} (h^{3} Mpc^{-3})$} & \colhead{$M_{B}^{\rm med}$} & \colhead{$\bar{z}$} & \colhead{$\frac{N_{\rm Q}}{N_{\rm Q}+N_{\rm Sf}}$}
}
\startdata
z0.43 & $0.35 - 0.5$ & $3.48 \times 10^{5}$ & 1650 & $ 4.74 \times 10^{-3}$ & -20.53 & 0.43 & 1605 & $ 4.61 \times 10^{-3}$ & -20.26 & 0.43 & 0.51\\
z0.57 & $0.5 - 0.65$  & $5.42 \times 10^{5}$ & 1818 & $ 3.35 \times 10^{-3}$ & -20.77 & 0.58 & 3258 & $ 6.01 \times 10^{-3}$ & -20.35 & 0.57 & 0.36\\
z0.73 & $0.65 - 0.8$  & $7.33 \times 10^{5}$ & 2291 & $ 3.12 \times 10^{-3}$ & -20.87 & 0.73 & 4570 & $ 6.23 \times 10^{-3}$ & -20.56 & 0.73 & 0.33\\
z0.88 & $0.8 - 0.95$  & $9.09 \times 10^{5}$ & 2509 & $ 2.75 \times 10^{-3}$ & -21.06 & 0.87 & 6002 & $ 6.6 \times 10^{-3}$ & -20.82 & 0.88 & 0.29\\
z1.00 & $0.95 - 1.1$  & $1.06 \times 10^{6}$ & 2182 & $ 2.05 \times 10^{-3}$ & -20.91 & 1.02 & 7768 & $ 7.30 \times 10^{-3}$ & -20.74 & 1.03 & 0.22\\
\enddata
\tablecomments{$V$ is the volume covered by ALHAMBRA in each redshift bin. For each of the samples selected by spectral type we show the number of galaxies $N$, the mean number density $\bar{n}$, the median $B$-band absolute magnitude $M_B^{\rm med}$ and the mean redshift $\bar{z}$. The last column gives the fraction of early-type galaxies in the bin. NW ALH-4 frame is not included.}
\end{deluxetable*}

\begin{figure}
\begin{center}
\includegraphics[height=0.35\textheight]{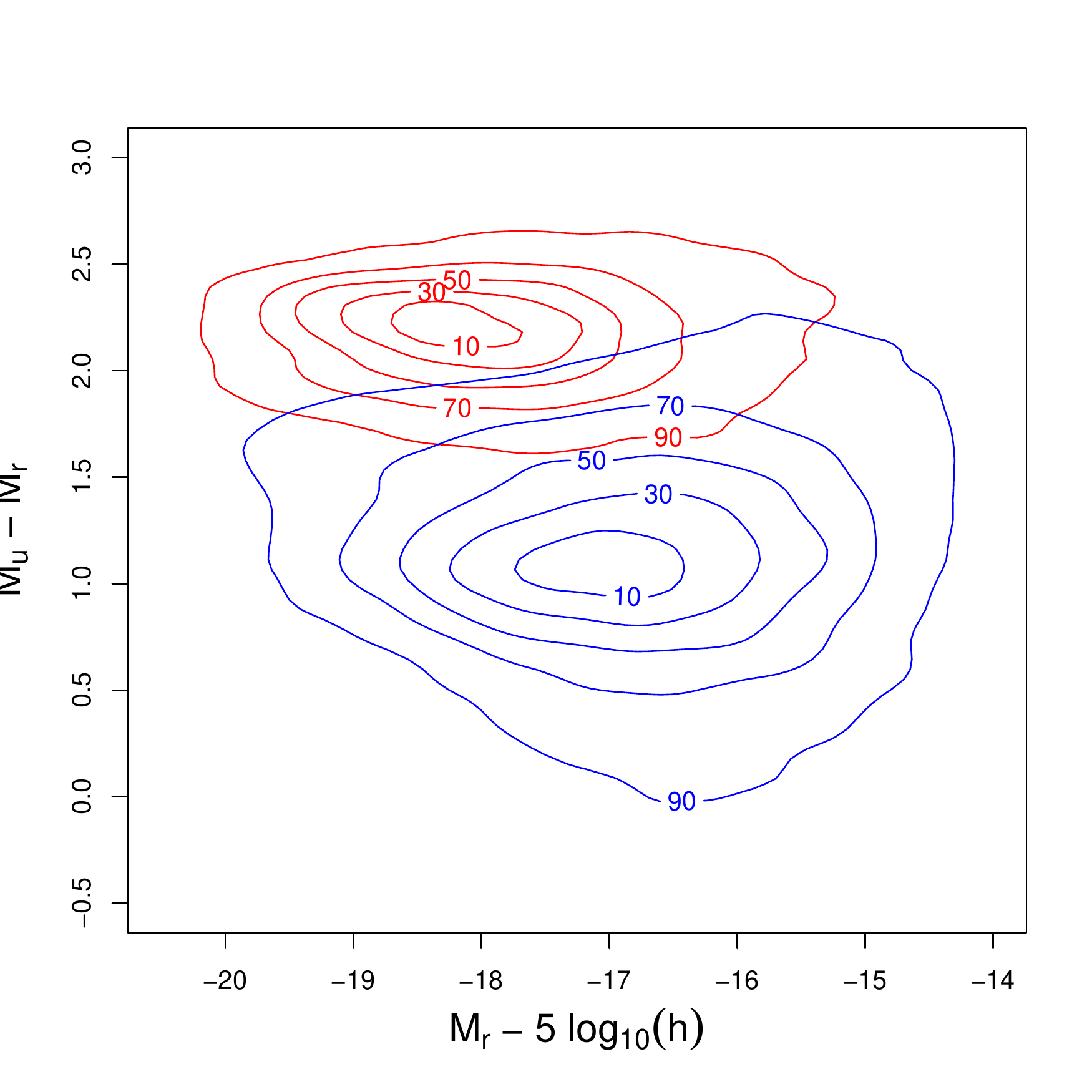}\\ 
\end{center}
\caption{\label{color}
Absolute rest-frame broad-band colour-magnitude diagram for galaxies
with redshifts between 0.35 and 0.75. Quiescent and star-forming galaxies
(selected by their best-fit spectral type) are shown, respectively, as
red and blue percentile contours. We have used SDSS absolute magnitudes derived from
the ALHAMBRA photometry as described in the text. Our classification
by (photometric) spectral type closely matches the usual broad-band
colour selection.}
\end{figure}

In Fig.~\ref{color} we show how our classification of quiescent and
star-forming galaxies performs on a colour-luminosity diagram. We plot
$M_r$ and $M_u$, which correspond to the absolute magnitudes in the
SDSS rest-frame broad-band filters $r$ and $u$, and were estimated
from ALHAMBRA data by \cite{2011Stefanon} for galaxies with redshift
$0.35 < z < 0.75$ and good quality photometric redshifts. We see how
well the ALHAMBRA spectral-type classification reproduces the expected
behaviour \citep{2004ApJ...608..752B}: quiescent galaxies correspond
to the `red sequence' in the diagram, while star-forming galaxies form
the `blue cloud'.  In addition to the clear segregation in colour, we
see that quiescent galaxies show, on average, slightly brighter
luminosities than star-forming ones. This shows that our selection by
(photometric) spectral type is almost equivalent to a selection in
broad-band colour.

\begin{figure*}
\begin{center}
\includegraphics[height=0.47\textheight]{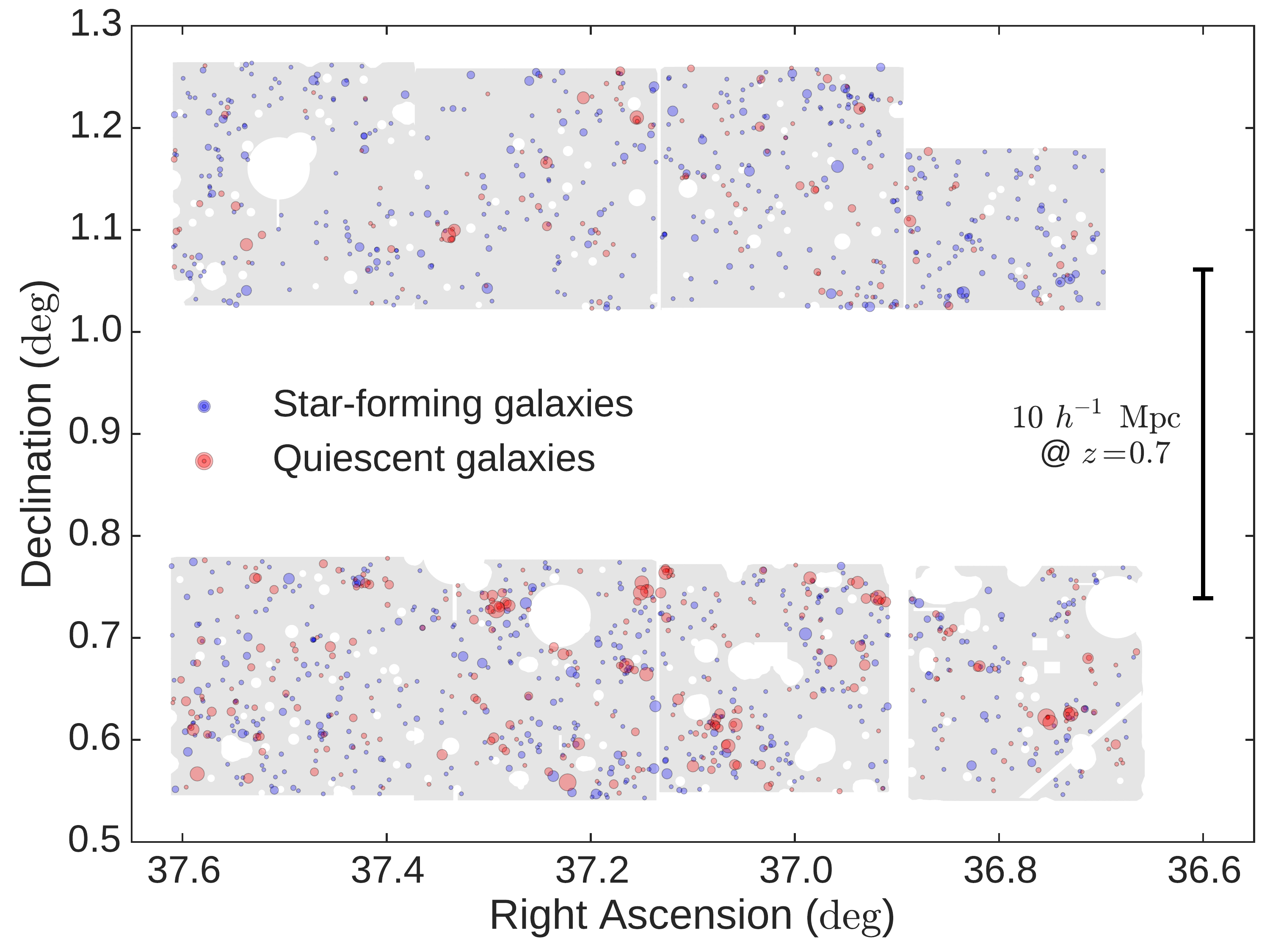}\\
\includegraphics[height=0.47\textheight]{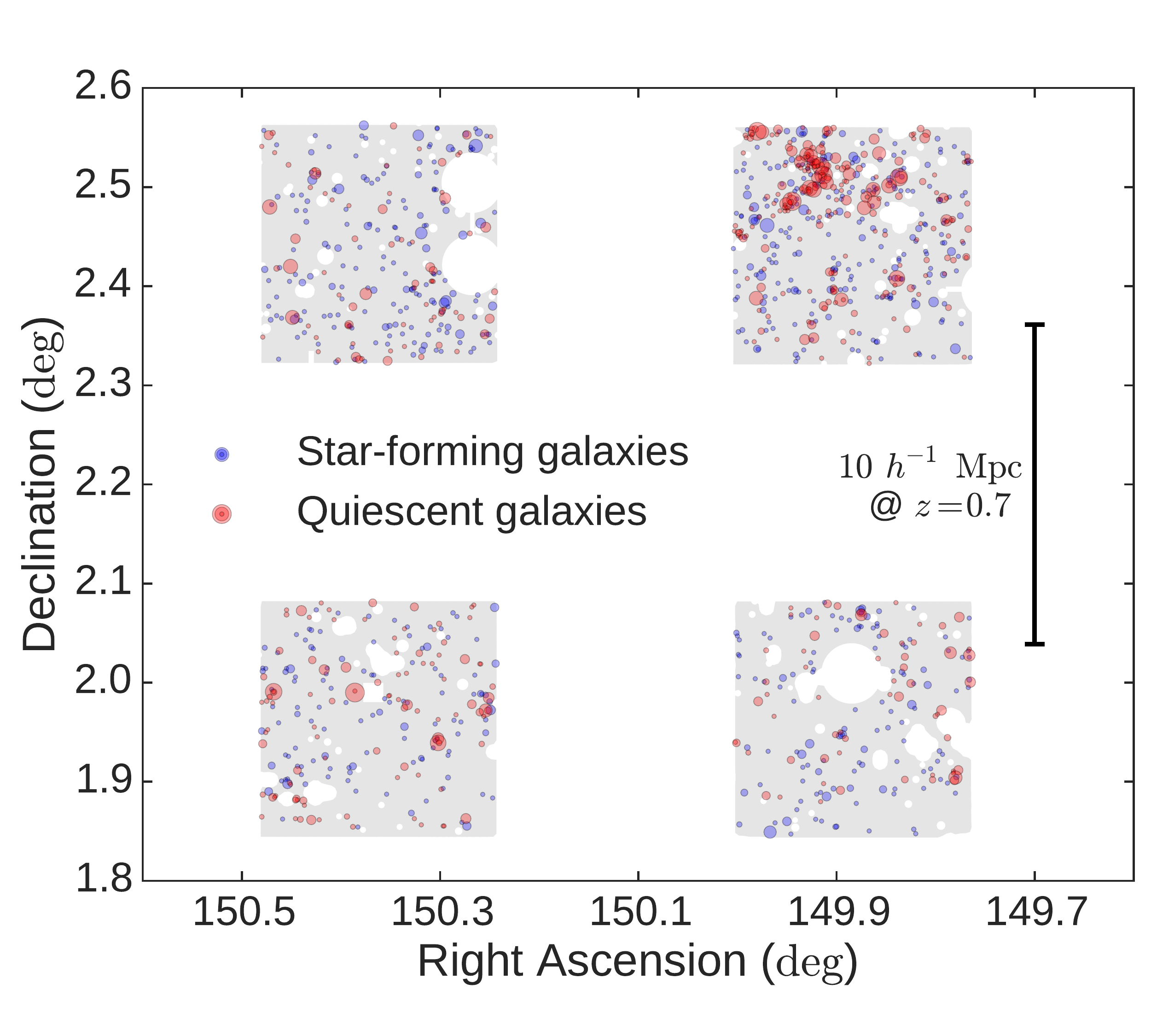}\\
\caption{\label{fields}Projection onto the sky of the z0.73 sample ($0.65 < z < 0.8$) for two ALHAMBRA fields: ALH-2 (top) and ALH-4/COSMOS (bottom). Galaxies have been coloured according to their type: blue circles correspond to star-forming and red circles to quiescent galaxies, and the size of each circle is proportional to the luminosity of the corresponding galaxy. North is to the top and East is to the left. The diagram shows the geometry of the ALHAMBRA fields, with the angular mask described in the text displayed as a light-grey background. The scale of 10 $h^{-1}$ Mpc at $z=0.7$ is indicated as a vertical bar. A heavy concentration of red circles (quiescent galaxies), corresponding to the big coherent structure described in the text, is patent in the NW quadrant of the ALH-4/COSMOS field.}
\end{center}
\end{figure*}

In Fig.~\ref{fields} we show the projection of two fields, ALH-2 and ALH-4/COSMOS, onto the plane of the sky. The coherent superstructure in the ALH-4/COSMOS field at $0.6 < z < 0.8$ is well appreciated. The skeleton of the structures, that form the cosmic web, is perfectly delineated by the red quiescent galaxies, while blue star-forming ones tend to populate the field or lower density regions. A similar trend was also visible in the redshift versus right ascension diagram of \citet{2013Msngr.151...41G}. We will further study this colour-density relation \citep{2006A&A...458...39C} in the following sections by means of the projected correlation function.

Finally, we remove the North-West ALH-4 frame from the analysis (the top-right section on the bottom panel of Fig.~\ref{fields}). As seen in \citetalias{2014MNRAS.441.1783A}, there exists an anomalous clustering in the ALH-4 field, which overlaps with the Cosmic Evolution Survey \citep[COSMOS,][]{Scoville2007h}.
It is well known that the COSMOS survey presents higher clustering amplitude than similar surveys \citep[see, e.g.,][]{McCracken2007a,tor10a}, due to the presence of large overdense structures in the field \citep{Guzzo2007a,Scoville2007d}.
This overdensity of structures is also observed in ALHAMBRA when comparing this peculiar region of the ALH-4 field with the rest of the fields \citep{2013arXiv1306.4968M,Ascaso2015a}. In \citetalias{2014MNRAS.441.1783A} the authors showed that ALH-4/COSMOS is an outlier in terms of clustering. We have seen that not only does this region introduce anomalies in the measurement of the clustering statistics, but it also affects the error estimation of these statistics.
In \citetalias{2014MNRAS.441.1783A} the authors also identified ALH-7/ELAIS-N1 as an outlier field (although the significance of the anomaly was smaller in this case).
However, here we do not find any significant change in our results when removing the ALH-7/ELAIS-N1 field, so we keep it in for all our calculations.

For each redshift bin we will analyse the clustering for the full 
selected population and separately for quiescent and star-forming galaxies.  Table~\ref{population} summarises the different samples
in each redshift bin.  Columns $N_{\rm Q}$ and $N_{\rm Sf}$ are the
number of quiescent and star-forming galaxies respectively, and the last
column is the fraction of quiescent galaxies in each of the redshift
subsamples.

We see that the fraction of quiescent galaxies decreases with
redshift. This behaviour is expected qualitatively, as blue
star-forming galaxies are dominant at earlier cosmic times, while red
quiescent galaxies appear late, once star formation
stops. This trend was also observed in a similar redshift range by
e.g. \cite{2009A&A...508.1217Z} for the zCOSMOS 10k bright sample. They found that
the population of bright late-type galaxies becomes dominant at higher redshifts, and therefore the fraction of early-type galaxies decreases with redshift accordingly.

\section{Methods}
\label{sec:methods}

The method used for the correlation analysis of our data follows closely the one used in \citetalias{2014MNRAS.441.1783A}, where it is discussed in detail.
We present here a summary of the methods and some points where the details differ. 
We estimate the correlation function using the estimator proposed by \cite{1993ApJ...412...64L} and the projected correlation function to recover real-space clustering from our photometric redshift catalogues as described by \cite{1983ApJ...267..465D} and \cite{2009MNRAS.394.1631A}. 
We use the delete-one jackknife method for the error estimation and linear regression to fit the different model correlation functions to our data.

\subsection{Estimation of the projected correlation function}
\label{sec:estimation-wp}

The method introduced by \citeauthor{1983ApJ...267..465D} is based on the decomposition of pair separations in distances parallel and perpendicular to the line-of-sight, $(r_{\parallel},r_p)$.  
Given two galaxies, if we define their radial vectors to the observer as $\mathbf{s_1}$ and $\mathbf{s_2}$, then their separation vector is $\mathbf{s} \equiv \mathbf{s_1} - \mathbf{s_2}$ and the line-of-sight vector is $\mathbf{l} \equiv \mathbf{s_1} + \mathbf{s_2}$. 
From these, we can now calculate the transverse and radial distances as
\begin{equation}\label{distv}
r_{\parallel} \equiv \frac{|\mathbf{s} \cdot \mathbf{l}|}{|\mathbf{l}|}, r_{p} \equiv \sqrt{\mathbf{s} \cdot \mathbf{s} - r_{\parallel}^2} \, .
\end{equation}
Once $(r_{\parallel},r_p)$ are defined for each galaxy pair we can proceed to calculate the two-dimensional correlation function, $\xi(r_{\parallel},r_p)$ in an analogous way to $\xi(r)$ \citep{2002sgd..book.....M}. 
With this method, for every galaxy pair we define a plane passing through the observer and containing the vectors $\mathbf{s_1}$ and $\mathbf{s_2}$. 
Therefore, we only need to assume isotropy in the plane perpendicular to the line of sight. 

To estimate the projected correlation function we need a random Poisson catalogue with the same selection function as our data. 
We create this auxiliary catalogue in each case using the software \textsc{Mangle} \citep{ham04a,swa08a} to apply the angular selection mask defined in \citetalias{2014MNRAS.441.1783A}.
In addition, the random points are distributed so that they follow the redshift distribution of each of the galaxy samples used (see Section \ref{sec:seg}).
We estimate the two-dimensional two-point correlation function as \citep{1993ApJ...412...64L}
\begin{equation}\label{LS}
\hat{\xi}(r_{\parallel},r_p) = 1 + {\left(\frac{N_R}{N_D}\right)}^2\frac{DD(r_{\parallel},r_p)}{RR(r_{\parallel},r_p)} - 2\frac{N_R}{N_D}\frac{DR(r_{\parallel},r_p)}{RR(r_{\parallel},r_p)} \, ,
\end{equation}
where $DD(r_{\parallel},r_p)$, $DR(r_{\parallel},r_p)$ and $RR(r_{\parallel},r_p)$ correspond to pairs of points with transverse separations in the interval $[r_p, r_p + \mathrm{d}r_p]$ and radial separations in the interval $[r_\parallel, r_\parallel + \mathrm{d}r_\parallel]$. $DD$ counts pairs of points in the data catalogue, $RR$ counts pairs in the random Poisson catalogue, and $DR$ counts crossed pairs between a point in the data catalogue and a point in the Poisson catalogue. 
$N_D$ is the number of points in the data catalogue, and $N_R$ ($= 20 N_D$) is the number of points used in our random Poisson catalogue.

We can define the projected correlation function as
\begin{equation}\label{int}
w_p(r_p) = 2\int^{\infty}_{0}\hat{\xi}(r_{\parallel},r_p) \mathrm{d}r_{\parallel} \, .
\end{equation}
As $w_p$ depends only on $r_p$, and the angle between any pair of points is small, it will not be significantly affected by redshift errors, as these will mainly produce shifts in $r_{\parallel}$. 
For computational reasons, we have to fix a finite upper limit, $r_{\parallel, \rm max}$, for the integral in Eq.~\ref{int}. The authors showed in \citetalias{2014MNRAS.441.1783A} that the optimal value for our samples is  $r_{\parallel,\rm max} = 200\, h^{-1}$~Mpc.

We further correct our measured $w_p(r_p)$ for the bias introduced by the integral constraint \citep{1980lssu.book.....P}.
This effect arises because we are measuring the correlation function with respect to the mean density of a sample instead of the global mean of the parent population.
We base our correction on the effect of the integral constraint on the three-dimensional correlation function $\xi(r)$. 
This function is biased to first order as
\begin{equation}
\xi(r) = \xi^{\rm true}(r) - K \, ,
\end{equation}
where $K$ is the integral constraint term. Using eq.~(\ref{int}), the effect on $w_p(r_p)$ is then
\begin{equation}
\label{eq:wpKcorrect}
w_p^{\rm true}(r_p, r_{\parallel,  \rm max}) = w_p(r_p, r_{\parallel, \rm max}) + 2Kr_{\parallel, \rm max} \, .
\end{equation}
Given a model correlation function, $K$ can be estimated as \citep{Roche1999o}
\begin{equation}
\label{eq:Kest}
K \simeq \frac{\sum_i RR(r_i)\xi^{\rm model}(r_i)}{\sum_i RR(r_i)} = \frac{\sum_i RR(r_i)\xi^{\rm model}(r_i)}{N_R(N_R-1)} \, .
\end{equation}
We proceed in an iterative way to introduce this correction for each of our samples.
We first fit our original $w_p(r_p)$ measurements to a double power-law model, $A\cdot x^\beta + C\cdot x^\delta$. We use the model $\xi(r)$ obtained from this fit to estimate $K$ using Eq.~\ref{eq:Kest}, and obtain our corrected values of $w_p(r_p)$ from Eq.~(\ref{eq:wpKcorrect}).
We use the corrected values to perform the model fits described in Sects.~\ref{sec:power} and \ref{sec:depend-bias-spectr}, and for all the results reported in Sect.~\ref{sec:res}.
In any case, the effect of the integral constraint in our measurements is always much smaller than the statistical errors.

To estimate the correlation function errors for each bin in $r_p$, we used the jackknife method \citep[see, e.g.,][]{nor08a}. 
We divided our volume in $N_\textrm{jack} = 47$ equal sub-volumes, corresponding to the individual ALHAMBRA frames \citep[see][]{2013arXiv1306.4968M}, and constructed our jackknife samples omitting one sub-volume at a time. 
We repeated the full calculation of $w_p(r_p)$ (including the integral constraint correction) for each of these samples. 
Denoting by $w_{pi}^{k}$ the correlation function obtained for bin $i$ in the jackknife sample $k$, the covariance matrix of the projected correlation function is then 
\begin{equation}
\label{eq:covmat}
\Sigma_{ij} = \frac{N_{\rm jack} - 1}{N_{\rm jack}}\sum_{k=1}^{N_{\rm jack}}(w_{p}^{k}(r_i)-\bar{w}_{p}(r_i)) \cdot (w_{p}^{k}(r_j)-\bar{w}_{p}(r_j)) \, ,
\end{equation}
where $\bar{w}_{pi}$ is the average of the values obtained for bin $i$. 
The errors for individual data points (shown as errorbar in the plots) are obtained from the diagonal terms of the covariance matrix as
\begin{equation}
\sigma_i = \sqrt{\Sigma_{ii}}.
\end{equation}

\section{Results and discussion}
\label{sec:res} 

\begin{figure}
\begin{center}
\includegraphics[height=0.5\textheight]{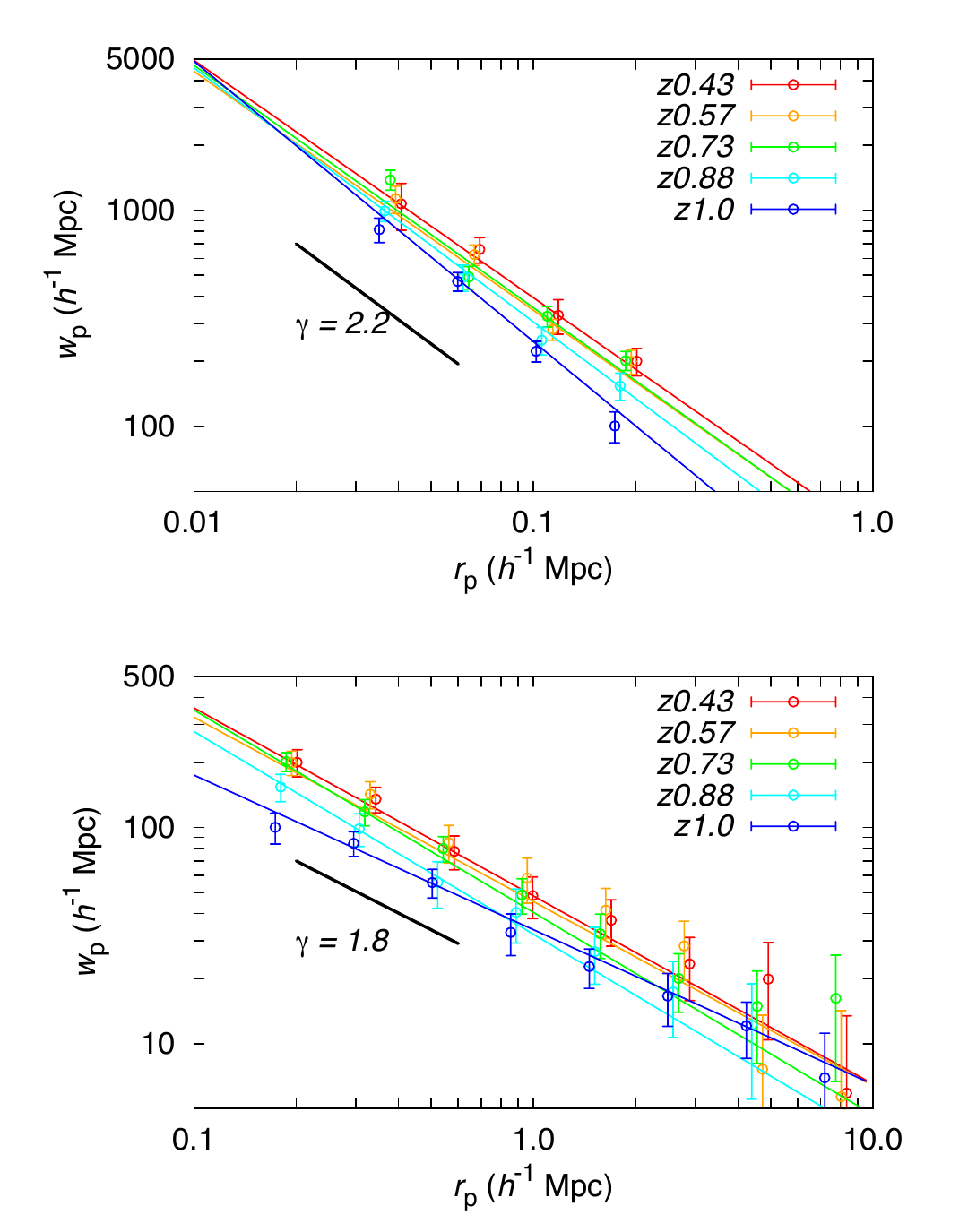}
\end{center}
\caption{\label{full}
Projected correlation function for the full population sample in each of the redshift bins (points with errorbars). Top: small scales ($0.03 < r_p < 0.2$). Bottom: large scales ($0.2 < r_p < 10.0$). Error bars are calculated with the delete-one jackknife method and values at the same $r_p$ are shifted for clarity.
Solid lines with matching colors show the best-fit power law in each case. The black segment represents the mean slope of the curves.}
\end{figure}

In this section we first present the results of the calculation of the projected correlation function $w_p(r_p)$ for the different samples described in Section~\ref{sec:seg}. This is done in Subsection~\ref{sec:power}. We also present the analysis of the bias (Subsection~\ref{sec:depend-bias-spectr}). The calculation has been performed for scales from $0.03$ to $10.0 \, h^{-1} $ Mpc for the projected correlation function and from $1.0$ to $10.0 \, h^{-1} $ Mpc for the bias. Fig.~\ref{full} shows the projected correlation function for the full samples. The first remarkable result that deserves to be pointed out is a clear change of the slope of the $w_p(r_p)$ functions around $r_p \sim 0.2 \, h^{-1} $ Mpc, as already mentioned by \cite{2006ApJ...644..671C} and \cite{coi08a}. 

In this section, we compare our results with previous works that studied the galaxy clustering and  its dependence on spectral type or colour in the redshift range $z \in [0,1]$, as mentioned in the introduction. 
Given the luminosity selection of our sample (see Section~\ref{sec:seg}), in each case we use for comparison the published results for volume-limited samples with number density closest to $n = 10^{-2} \, h^3 \, \mathrm{Mpc}^{-3}$.
The number density of the samples shown in our comparisons are within $20\%$ of this figure with two exceptions: the PRIMUS sample at $z \simeq 0.4$ (with number density of $n = 1.6 \times 10^{-2} \, h^3 \, \mathrm{Mpc}^{-3}$, \citealp{Skibba2013c}), and the VIPERS sample at $z \simeq 0.6$ (with number density of $n = 0.33 \times 10^{-2} \, h^3 \, \mathrm{Mpc}^{-3}$, \citealp{2013Marulli}).
In the case of \citet{Meneux2006a}, they use a flux-limited sample resulting in a evolving number density with redshift in the range $n = 0.33 - 1.2 \times 10^{-2} \, h^3 \, \mathrm{Mpc}^{-3}$.

\subsection{Power-law modelling}
\label{sec:power}

Power laws are simple and widely used models to describe the correlation function of the galaxy distributions, as they provide a very good approximation over a large range of scales with only two free parameters. The observed change of the slope mentioned above  forced us to model the projected correlation function $w_p$ by means of two power laws, one that fits the function at small scales and the other one at large scales. A similar treatment was done by \cite{2006ApJ...644..671C} in their analysis of the clustering in the DEEP2 survey at $z=1$. The departure from power-law behavior at small scales can be explained naturally in the framework of the halo occupation distribution (HOD) model that considers the contribution to the correlation function of pairs within the same halo (one-halo term), which is dominant at short scales, and the transition to the regime where the function is dominated by pairs from different halos (two-halo term), at large scales. We will present HOD fits to the ALHAMBRA data in a separate paper. Therefore, we fit two power laws as:

\begin{equation}
w_{p}^{s}(r_p) = A r_p^{\beta}, \text{if } r_p \leq r_s
\end{equation}
for the small scales, and

\begin{equation}
w_{p}^{l}(r_p) = C r_p^{\delta}, \text{if } r_p \geq r_s
\end{equation}
for the large ones. We fix value $r_s \simeq0.2 h^{-1}$ Mpc. An abrupt change in the projected correlation function has also been detected at this scale by \cite{2006A&A...457..145P} for the blue galaxies of the COMBO-17 sample. $A$, $\beta$, $C$ and $\delta$ are the free parameters. We treat each power law independently and express them in terms of the equivalent model for the three-dimensional correlation function $\xi$. 
\begin{equation}
\xi^{\rm pl}(r) = \left(\frac{r}{r_0}\right)^{-\gamma} \, .
\end{equation}
$A$ and $\beta$ (analogously, $C$ and $\delta$) can be related to the parameters $\gamma$ (power-law index) and $r_0$ (correlation length) as shown in \cite{1983ApJ...267..465D}: 
\begin{equation}
A = r_{0}^{\gamma} \frac{\Gamma(0.5)\Gamma\left[0.5(\gamma-1)\right]}{\Gamma(0.5\gamma)} \; , \; \beta = 1 - \gamma \, .
\end{equation}

We have performed the fitting of this model to our data using a standard $\chi^2$ method, by minimizing the quantity
\begin{equation}
\label{chim}
\chi^2(r_0, \gamma) = \sum_{i=1}^{N_{bins}} \sum_{j=1}^{N_{bins}} (w_p(r_i) - w_p^{pw}(r_i)) \cdot \Sigma_{ij}^{-1} \cdot (w_p(r_j) - w_p^{pw}(r_j)) \, ,
\end{equation}
where $\Sigma$ is the covariance matrix. We fit this model to our data at scales $0.03 \leq r_p \leq 0.2 h^{-1}$ Mpc and $ 0.2 \leq r_p \leq 10.0 h^{-1}$ Mpc for each sample using the covariance matrix computed from eq.~(\ref{eq:covmat}), to obtain the best-fit values of $r_0$, $\gamma$ and their uncertainties (see Table~\ref{fittablem}). This fitting has been performed using the POWERFIT code developed by \citet{2012ApJ...745..180M}.

We must remark that the statistical errors of the correlation function at different separations $r_p$ are heavily correlated because a given large-scale structure adds pairs at many different distances. The higher the values of the off-diagonal terms of the covariance matrix, the stronger the correlations between the errors. When this happens the best fit parameters $r_0$ and $\gamma$ might be affected as has been illustrated by \cite{2004ApJ...608...16Z} for the SDSS survey. One could ignore the error correlations and use only the diagonal terms, but this is not justified if these terms are dominant. The parameters of the fits are listed in Table~\ref{fittablem}.

\subsubsection{Full samples}

Fig.~\ref{full} (top panel) shows the measurements of the projected correlation function $w_p(r_p)$ for the full samples at the small scales ($0.03 \leq r_p \leq 0.2 h^{-1} $ Mpc). The bottom panel shows the same function for large scales ($0.2 \leq r_p \leq 10.0 h^{-1} $ Mpc). Looking at both diagrams, we confirm the rise of the correlation function  at small scales already detected by \citet{coi08a} with values of $\gamma \sim 2.2$ (for the slope of the three-dimensional correlation function). We can also appreciate in the top panel of Fig.~\ref{full} that the correlation functions are steeper for the high-redshift samples with values of $\gamma$ increasing from $\sim 2.1$ for the closest redshift bin ($z \sim 0.4$) to $\sim 2.3$ for the farthest ($z\sim 1$). The correlation length significantly decreases with increasing redshift (see also Table~\ref{fittablem}).

\begin{figure}
\begin{center}
\includegraphics[height=0.27\textheight]{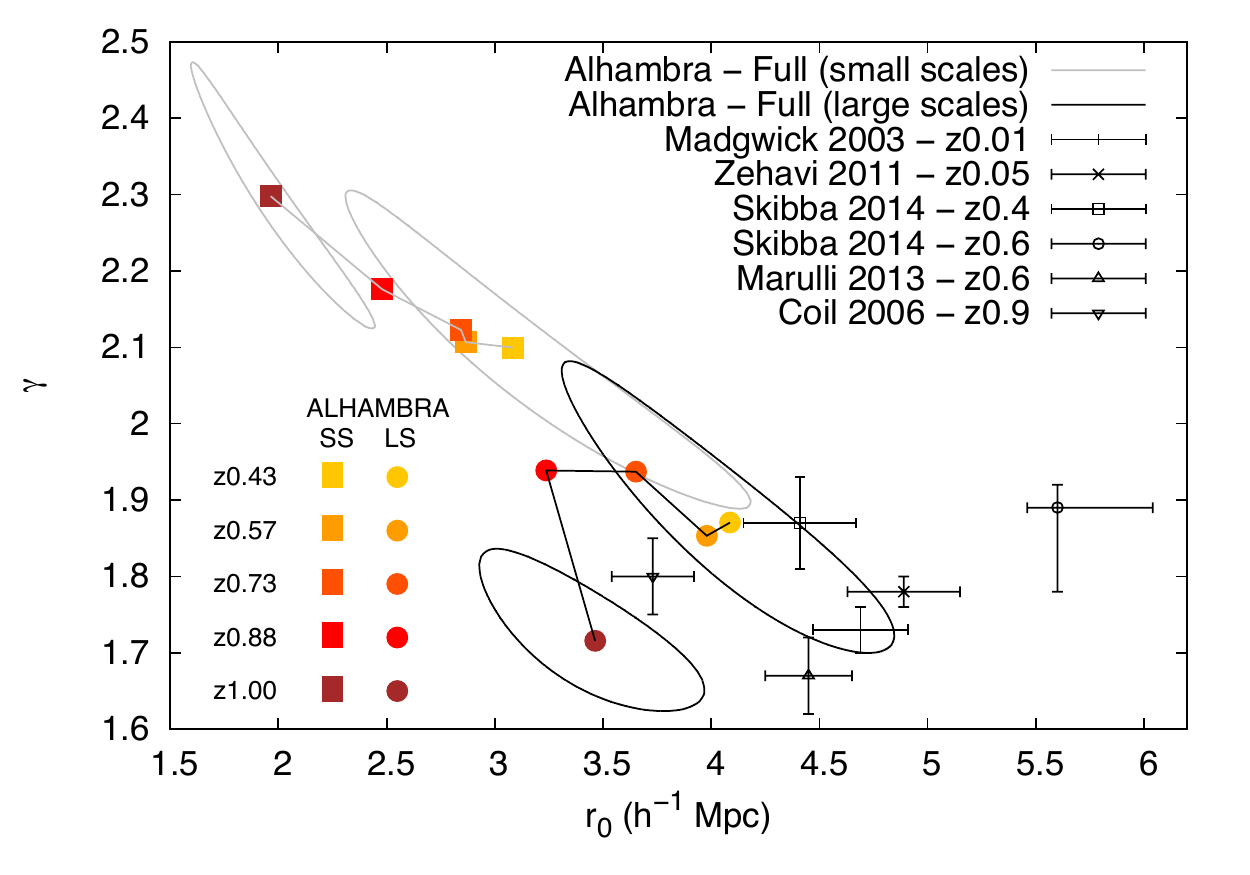}
\end{center}
\caption{\label{eli_full}
Parameters $r_0$, $\gamma$ obtained from the power-law fit to the projected correlation functions of our full population samples. In black, the $1\sigma$ confidence regions of the large scales fit ($0.2 < r_p < 10.0 h^{-1}$ Mpc) and in grey, the $1\sigma$ confidence regions of the small scales fit ($0.03 < r_p < 0.2 h^{-1}$ Mpc). For clarity, we show only the regions for the first and last redshift bin. Lines link the best-fit results for each sample accross different redshift bins.
For comparison, we show as points with errorbars the results of \cite{2003MNRAS.344..847M} (2dF), \cite{2011ApJ...736...59Z} (SDSS), \cite{2006ApJ...644..671C} (DEEP2), \cite{2013Marulli} (VIPERS) and \cite{Skibba2013c} (PRIMUS) (see the text for details). The parameters and their 1-sigma variation have been calculated using the method described in Sect.~\ref{sec:res}.} 
\end{figure}

For the large scales $ 0.2 \leq r_p \leq 10.0 h^{-1}$ Mpc, the slope of the correlation function is rather constant for all samples with values around $\gamma =1.8$, while again the correlation length decreases with redshift from $r_0 =4.1 \pm 0.5$ for $z\sim 0.4$ to $r_0=3.5 \pm 0.3$ for $z\sim 1$. 
The evolution of the amplitude indicates that the change in clustering is mainly driven by the overall growth of structure in the matter density field.
As we use for the fits the scales $0.2 < r_p < 10.0 h^{-1}$ Mpc (the 2-halo term becoming important at scales $r_p > 1.0 \, h^{-1}$ Mpc) the fact that the slope $\gamma$ does not significantly change also implies that the 2-halo contribution for this population does not significantly change its profile over this redshift interval.
All these effects were studied in detail in \citetalias{2014MNRAS.441.1783A} and extended to samples with different luminosities (see e.g. their figure~7).
We have seen that this is only broken at shorter scales, where the curve presents slightly higher values.

\begin{deluxetable*}{l|ccccc} 
\tablewidth{18cm}
\tablecaption{\label{fittablem}Results of the different fits to $w(r_p)$: power law and bias models}
\tablehead{
\colhead{Sample} & \multicolumn{5}{c}{Full population}\\
\colhead{} & \colhead{$r^s_0$} & \colhead{$\gamma^s$} & \colhead{$r^l_0$} & \colhead{$\gamma^l$} & \colhead{$b$}}
\startdata
z0.43 & $3.1 \pm 0.6$ & $2.1 \pm 0.13 $ & $ 4.1 \pm 0.5$ & $1.87 \pm 0.12$ & $1.21 \pm 0.14 $\\
z0.57 & $2.9 \pm 0.4$ & $2.11 \pm 0.1 $ & $ 4 \pm 0.5$ & $1.85 \pm 0.1$ & $1.23 \pm 0.17 $\\
z0.73 & $2.8 \pm 0.5$ & $2.12 \pm 0.12 $ & $ 3.7 \pm 0.4$ & $1.94 \pm 0.1$ & $1.25 \pm 0.14 $\\
z0.88 & $2.5 \pm 0.4$ & $2.18 \pm 0.1 $ & $ 3.2 \pm 0.7$ & $1.94 \pm 0.15$ & $1.2 \pm 0.5 $\\
z1.00 & $2 \pm 0.3$ & $2.3 \pm 0.11 $ & $ 3.5 \pm 0.3$ & $1.72 \pm 0.06$ & $1.3 \pm 0.13 $\\
\hline
\colhead{} & \multicolumn{5}{c}{Quiescent galaxies}\\
\hline
z0.43 & $4 \pm 1.2$ & $2.11 \pm 0.17 $ & $ 4.9 \pm 0.7$ & $1.89 \pm 0.16$ & $1.26 \pm 0.19 $\\
z0.57 & $2.3 \pm 0.5$ & $2.62 \pm 0.18 $ & $ 5.4 \pm 0.8$ & $1.85 \pm 0.11$ & $1.8 \pm 0.2 $\\
z0.73 & $3.6 \pm 0.7$ & $2.29 \pm 0.14 $ & $ 4.3 \pm 0.7$ & $2.15 \pm 0.16$ & $1.4 \pm 0.2 $\\
z0.88 & $4 \pm 0.9$ & $2.25 \pm 0.13 $ & $ 4.2 \pm 0.8$ & $2.14 \pm 0.17$ & $1.6 \pm 0.3 $\\
z1.00 & $3.5 \pm 0.9$ & $2.28 \pm 0.16 $ & $ 4.8 \pm 0.8$ & $1.8 \pm 0.13$ & $1.9 \pm 0.3 $\\
\hline
\colhead{} & \multicolumn{5}{c}{Star-forming galaxies}\\
\hline
z0.43 & $2.4 \pm 1.7$ & $2 \pm 0.5 $ & $ 4.3 \pm 0.5$ & $1.66 \pm 0.13$ & $1.33 \pm 0.18 $\\
z0.57 & $2.2 \pm 0.7$ & $2.1 \pm 0.2 $ & $ 3.6 \pm 0.4$ & $1.73 \pm 0.12$ & $1.21 \pm 0.17 $\\
z0.73 & $2.8 \pm 0.9$ & $2.1 \pm 0.2 $ & $ 3.5 \pm 0.4$ & $1.86 \pm 0.14$ & $1.18 \pm 0.14 $\\
z0.88 & $1.8 \pm 0.5$ & $2.3 \pm 0.2 $ & $ 3 \pm 0.4$ & $1.7 \pm 0.12$ & $1.2 \pm 0.4 $\\
z1.00 & $1.7 \pm 0.3$ & $2.34 \pm 0.13 $ & $ 3.2 \pm 0.3$ & $1.69 \pm 0.09$ & $1.25 \pm 0.13 $
\enddata
\tablecomments{Results of the fits of the power law model and the bias model to the data for each of our samples. $r_0^s$ and $\gamma^s$ correspond to the scales $0.03 < r_p < 0.2 h^{-1} $ Mpc, and $r_0^l$ and $\gamma^l$ to the scales $0.2 < r_p < 10.0 h^{-1} $ Mpc. These parameters have been calculated using the methods described in Sections~\ref{sec:power} and \ref{sec:depend-bias-spectr}}.
\end{deluxetable*}

The overall trend can be visualized in Fig.~\ref{eli_full}, where we show the evolution of the best-fit parameters of the three-dimensional correlation function $\xi(r)$ for small and large scales in the full population samples. Despite the great uncertainties, the diagram shows evolution with $r_0$ decreasing for both scale ranges as redshift grows. In addition, at small scales, the slope $\gamma$ also increases with redshift. The evolution, at large scales, of the correlation length extrapolates well to lower redshift with the value reported by \cite{2011ApJ...736...59Z} for the SDSS and by \cite{2003MNRAS.344..847M} for the 2dF galaxy redshift survey. \cite{2011ApJ...736...59Z} analysed the SDSS Main catalogue by means of the projected correlation function. They obtained values for the parameters $r_0$ and $\gamma$ by the same method used here, over the scale range $0.1 < r_p < 50 h^{-1}$ Mpc.
The values correspond to the galaxies selected in the luminosity bin $-20 < M_r < -19$ and $0.027 < z < 0.064$, with a number density ($n = 10.04 \times 10^{-3}\, h^3 \, \mathrm{Mpc}^{-3}$) and typical luminosity ($L^{\rm med}/L^{\star} = 0.4$) similar to the ALHAMBRA sample used in this work, so this is, qualitatively, a valid comparison. As we can see in Fig.~\ref{eli_full}, the slope of the correlation function for the full SSDS main sample is $\gamma =1.78 \pm 0.02$ compatible within one $\sigma$ with the values obtained for the ALHAMBRA survey at higher redshift within the range of large scales analysed here, and the correlation length $r_0 =4.89 \pm 0.26$ follows the evolutionary trend delineated by the ALHAMBRA higher redshift samples: $r_0$ increases at lower redshifts. Very similar results have been obtained by \cite{2003MNRAS.344..847M} for the 2dFGRS with  $\gamma =1.73 \pm 0.03$ and $r_0 =4.69 \pm 0.22$ within the range $ 0.2< r_p < 20.0 h^{-1}$ Mpc in the redshift interval $0.01< z < 0.015$. At larger redshift our results can be compared with the ones reported by \cite{Skibba2013c} for the PRIMUS survey. They have analysed two bins of redshift $ 0.2 < z < 0.5$ and $0.5 < z < 1$ with $M_g < -19$. In Fig.~\ref{eli_full} we have displayed their results for the correlation function parameters. We also plot a point corresponding to the VIMOS Public Extragalactic Redshift Survey (VIPERS) from \cite{2013Marulli} and another point corresponding to the DEEP2 survey from \cite{2006ApJ...644..671C}. All these results, for the three high redshift surveys, show perfect agreement with our own ALHAMBRA results.

It is important to understand the correlation between parameters $\gamma$ and $r_0$, as its interpretation can be delicate. 
If, for instance, $\gamma$ grows with the redshift of the sample, $r_0$ will tend to reduce its value, as $\xi(r)=1$ for shorter distances, as it can be appreciated in the top-left points (short $r_p$ scales) displayed in Fig.~\ref{eli_full} 
We must have this in mind for a proper understanding of our results. 
On the other hand, the decrease of $r_0$ with increasing redshift when $\gamma$ does not change, as we find in bottom-right points (large scales) in Fig.~\ref{eli_full}, can be interpreted as a self-similar growth of the structure at the calculated scales. 
This effect is specially reflected in the tilt of the confidence ellipses in Fig.~\ref{eli_full}, which shows the negative correlation between $r_0$ and $\gamma$.

\subsubsection{Segregated samples}

\begin{figure*}
\begin{center}
\includegraphics[height=0.75\textheight]{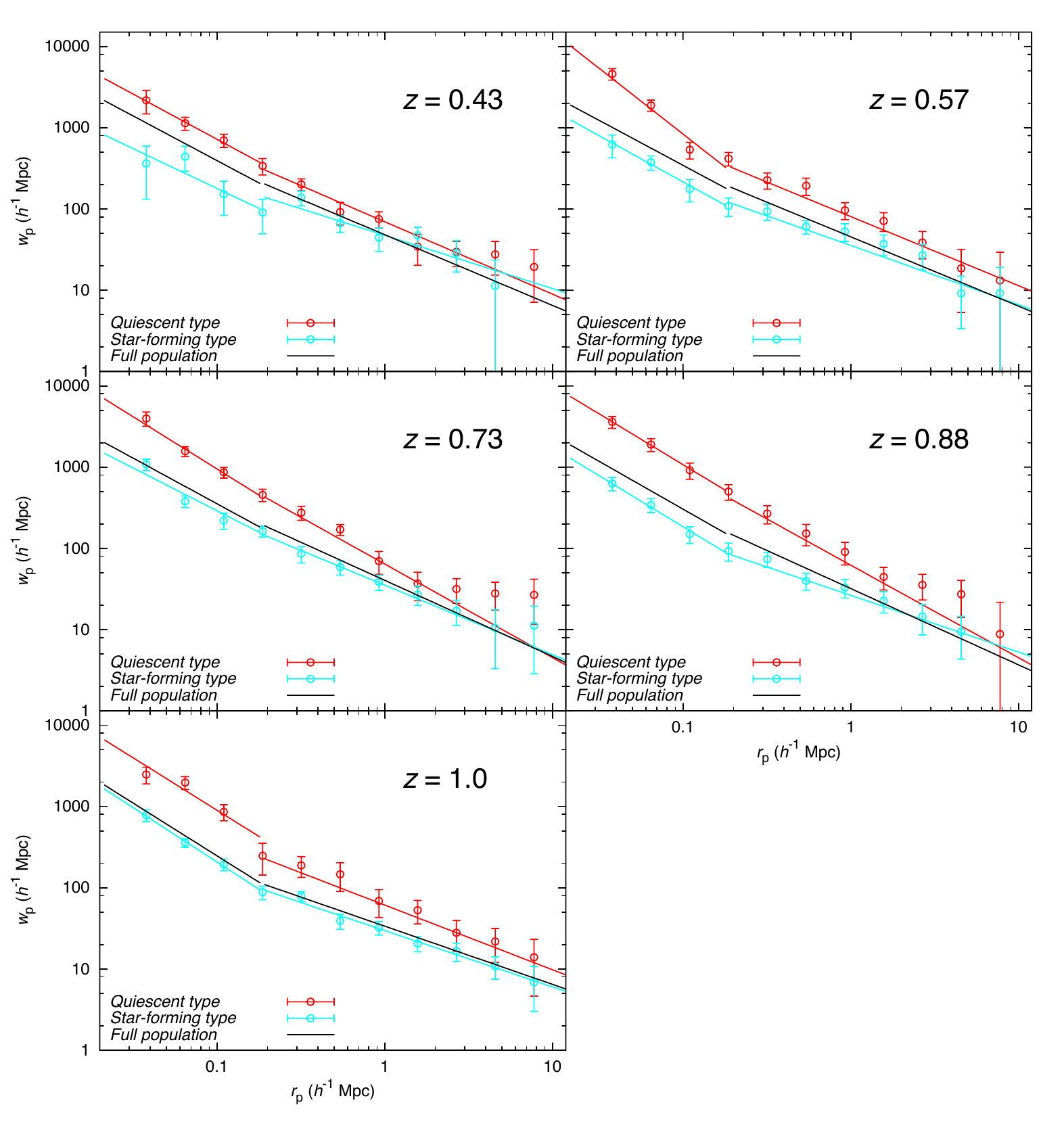}\\
\end{center}
\caption{\label{type}
Projected correlation functions for  quiescent (red) and star-forming (blue) galaxies (points with errorbars). 
Solid lines with matching colors show the best-fit power law in each case.
For reference, we also show the results for the full population with the continuous black line.
From top to bottom, left to right, the five reshift bins:  ($0.35 < z < 0.5$), ($0.5 < z < 0.65$), ($0.65 < z < 0.8$), ($0.8 < z < 0.95$) and ($0.95 < z < 1.1$). 
Error bars are calculated with the delete-one jackknife method.}
\end{figure*}

Fig.~\ref{type} shows the projected correlation function $w_p(r_p)$ for the quiescent and star-forming galaxies at the five redshift bins, compared to the full population.
As expected, the full population result occupies an intermediate position at low redshift, but evolves with redshift towards star-forming positions. This is expected due to the higher abundance of the latter in our samples, specially at high redshift. A visual inspection of Fig.~\ref{type} suggests that the projected correlation function shows the double slope corresponding to the 1-halo and the 2-halo terms, specially for the star-forming galaxies, due to their tendency to cluster in lower mass halos with smaller virial radii \citep{2000MNRAS.318..203S}.

Quiescent galaxies show a higher clustering at every redshift bin. In order to study the change of the clustering properties with redshift and spectral type, we fit the projected correlation function $w_p(r_p)$ of each sample with a power law model, using the method described above.

\begin{figure}
\begin{center}
\includegraphics[height=0.27\textheight]{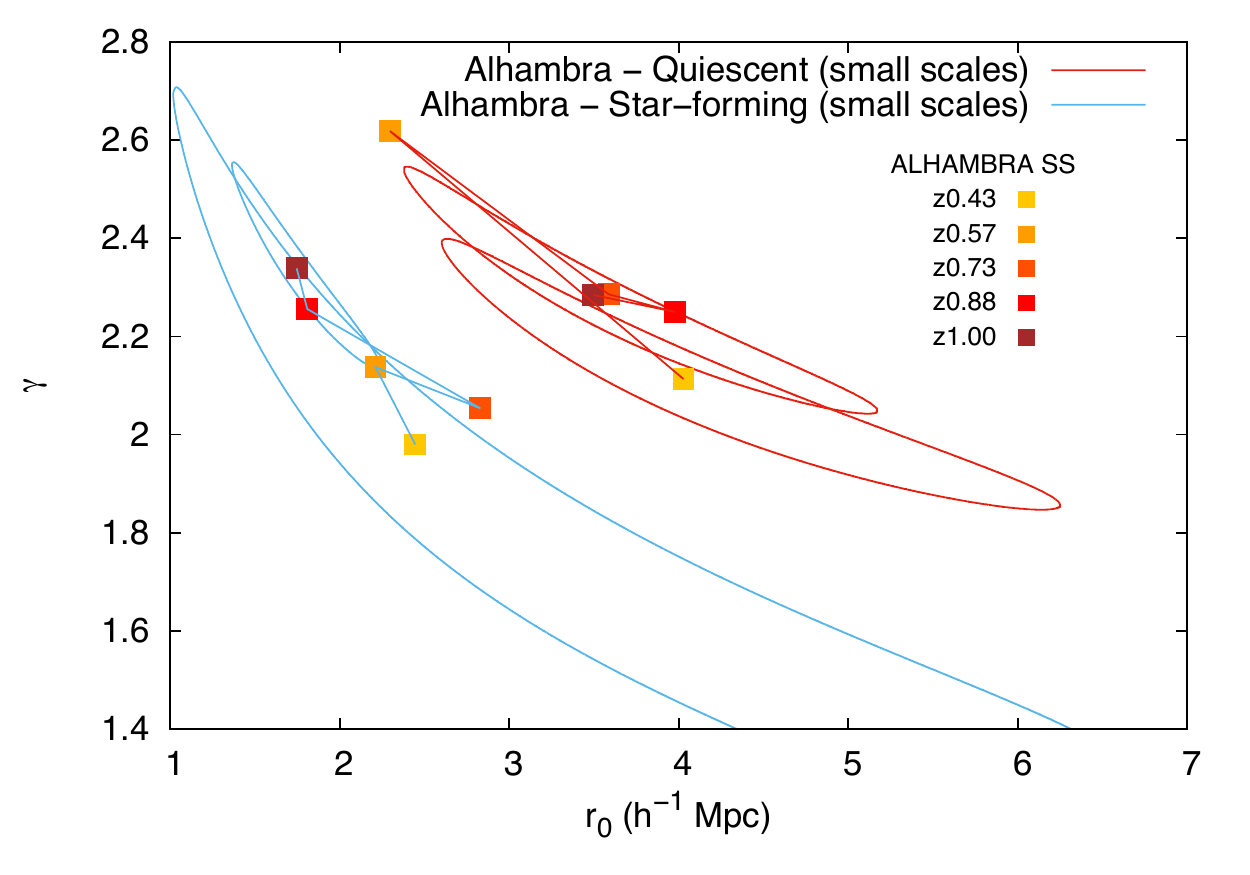}
\end{center}
\caption{\label{eli_seg_esq}
Parameters $r_0$, $\gamma$ obtained from the power-law fit to the projected correlation functions of our spectral segregated samples for the small scales ($r_p \leq 0.2 h^{-1}$). In red, the $1\sigma$ confidence regions of the quiescent galaxies fit and in blue, the $1\sigma$ confidence regions of the star-forming galaxies fit. For clarity, we show only the regions for the first and last redshift bin. Lines link the best-fit results for each sample accross different redshift bins.
The parameters and their 1-sigma variation have been calculated using the method described in Section~\ref{sec:power}}
\end{figure}

The amplitude of their correlation functions, as well as their slope, is higher than that for the star-forming galaxies in all cases. As for the full population we have modelled the correlation function with two different power laws at scales larger and smaller than $r_p = 0.2  h^{-1}$ Mpc. Star-forming galaxies show for all redshift bins a clear rise in their correlation function at small separations. As mentioned in the introduction, \cite{coi08a} found the same result for the bright blue galaxies of the DEEP2 galaxy redshift survey. They found that the effect is more pronounced at higher redshift corresponding to brighter galaxies. For the quiescent galaxies we would have fitted a single power law for the whole range, in particular for some redshift bins. However we have proceeded in the same way for the two galaxy types in order to simplify the analysis of the segregation. The comparison of the best-fit model to the data in each case is shown in Fig.~\ref{type}, and we see an excellent agreement in all cases. The parameters obtained from the fits are listed in Table~\ref{fittablem}.

\begin{figure}
\begin{center}
\includegraphics[height=0.27\textheight]{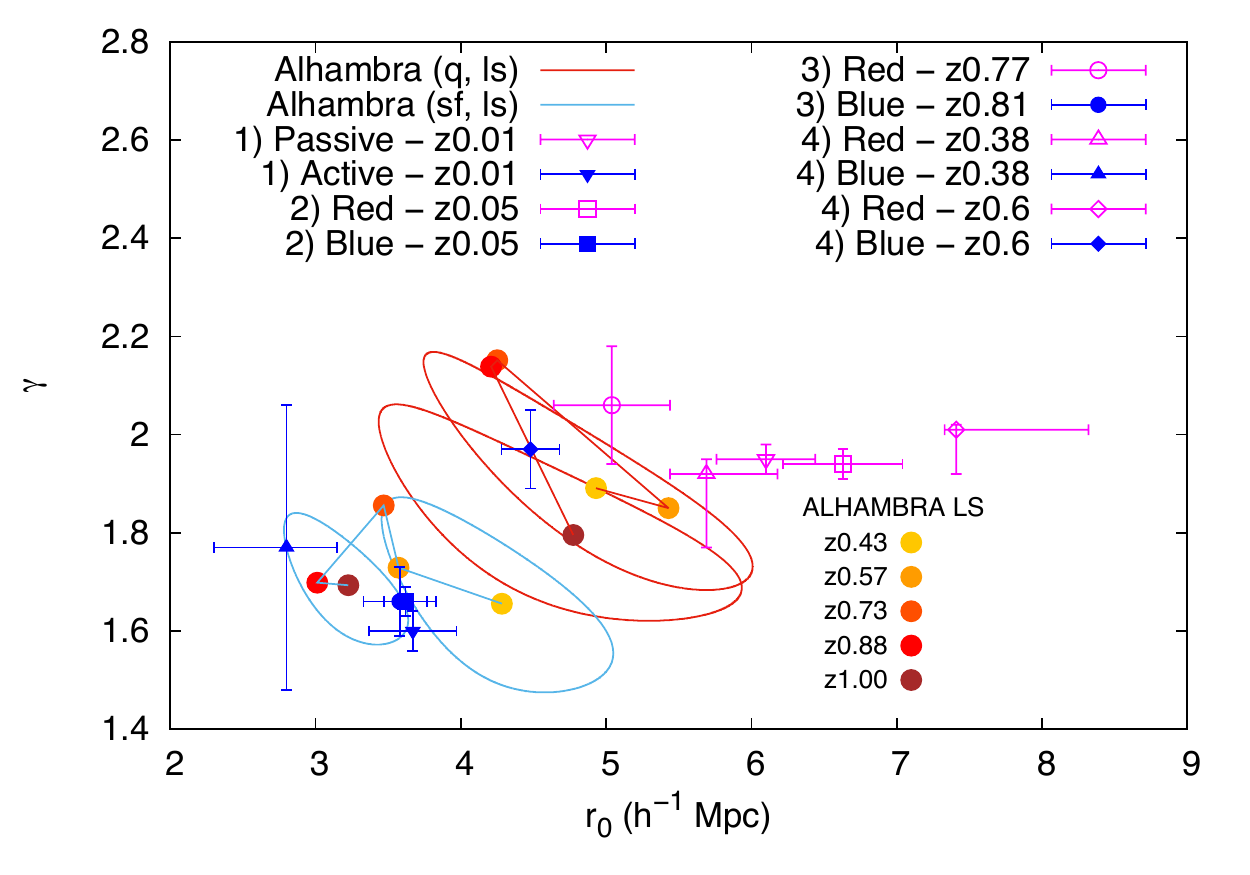}
\end{center}
\caption{\label{eli_seg_dre}
Parameters $r_0$, $\gamma$ obtained from the power-law fit to the projected correlation functions of our spectral segregated samples. In red, the $1\sigma$ confidence regions of the quiescent galaxies fit and in blue, the $1\sigma$ confidence regions of the star-forming galaxies fit. For clarity, we show only the regions for the first and last redshift bin. Lines link the best-fit results for each sample across different redshift bins.
For comparison, we show as points with errorbars the results of \cite{2011ApJ...736...59Z} and \cite{coi08a} (see the text for details). The parameters and their 1-sigma variation have been calculated using the method described in Section~\ref{sec:power}. For comparison, we plot the results obtained by 1) \cite{2003MNRAS.344..847M} (2dF) at $z \sim 0.01$, 2) \cite{2011ApJ...736...59Z} (SDSS) at $z \sim 0.05$, 3) \cite{coi08a} (DEEP2) at $z \sim 0.9$ and 4) \cite{Skibba2013c} (PRIMUS) at $z \sim 0.38$ and $z \sim 0.6$.}
\end{figure}

As we have done for the full population, to visualize if there is any evolution of the correlation function parameters we show the diagram of $\gamma$ vs. $r_0$ in Figs.~\ref{eli_seg_esq} and ~\ref{eli_seg_dre} for the segregated populations with the corresponding confidence regions, separated in the two scale regimes. In both cases (short and large scales) the parameter space occupied by quiescent galaxies can be clearly distinguished from the space occupied by star-forming galaxies, the first ones showing larger correlation length for both scaling ranges with the difference between both types well over 3$\sigma$ for small scales and about 2$\sigma$ for large scales.

At short scales the exponent of the correlation function $\gamma$ is similar for both galaxy types with values around $\gamma \sim 2.2$ (Fig.~\ref{eli_seg_esq}). The correlation length for star-forming galaxies varies roughly in the range $r_0= [2,3] \, h^{-1}$ Mpc. A visual hint of evolution could be appreciated for the star-forming galaxies, with steeper correlation functions (and lower correlation lengths) for higher redshifts, nevertheless, given the large error bars (see Table~\ref{fittablem}), this trend is not really significant. The parameters of the correlation function for quiescent galaxies at small scales do not show any evolution at all with values for $\gamma$ in the range $[2.1, 2.3]$ and correlation lengths in the range $r_0= [3.5,4.0] \, h^{-1}$ Mpc for all redshift bins except for the second bin ($z=0.57$), which displays a higher value of the exponent $\gamma$ and smaller $r_0$.  The ALHAMBRA survey has allowed us to measure the behaviour of the clustering properties of the segregated samples at these very short scales. These scales had not been previously studied with the detail that we are showing here because other samples cannot reliably estimate the correlation function at $r_p < 0.1  h^{-1}$ Mpc because they are not deep enough, or dense enough at these distance, due for example to fiber collisions in the case of spectroscopic surveys \citep{2012ApJ...756..127G}.

Fig.~\ref{eli_seg_dre} shows the same results at scales $0.2 < r_p < 10.0  h^{-1}$ Mpc. For this scale range, we can compare with the results from other authors. 
We see again that the regions of parameter space occupied in the diagram for quiescent and star-forming galaxies are different.
Star-forming galaxies present both lower exponent $\gamma$ and lower correlation length $r_0$ than quiescent galaxies. These differences are significant at the 2$\sigma$ level.
The value of $\gamma$, exponent of the correlation function, is roughly constant for all redshift bins and is $\sim 1.7$. 
A hint of evolution can be seen in the correlation length, since $r_0$ decreases from $r_0 \sim 4.3$ to $r_0 \sim 3 \, h^{-1}$ Mpc with increasing redshift, which corresponds to a $\sim2\sigma$ change in $r_0$.
The values of the correlation function parameters reported by other authors for different samples at lower and similar redshift are compatible with the ALHAMBRA results shown here.
In the diagram we see that our fits are consistent with the points corresponding to the correlation function parameters of the active galaxies from the 2dFGRS \citep{2003MNRAS.344..847M} at $z \sim 0.01$, a blue subsample drawn from the SDSS-main  \citep{2011ApJ...736...59Z} at $z \sim 0.05$, a blue population of the DEEP2 redshift survey \citep{coi08a} at $z \sim 0.9$, and the blue sample from the PRIMUS survey \citep{Skibba2013c} at $z \sim 0.4$.
This result also agrees with the qualitative behaviour of the evolution  of the correlation length reported by \citet{Meneux2006a} from the VVDS sample where they conclude that the clustering amplitude of the late-type star-forming galaxies remains roughly constant since $z\sim 1.5$, although they found a slight rise of this amplitude at their larger redshift bin $1.2 < z <2.0$. 
However, one should bear in mind that \cite{Meneux2006a} use a flux-limited sample, so this evolution may be affected by the change in luminosity of the samples.
The values of the correlation length reported by  \citet{Meneux2006a} are slightly smaller than the values calculated here for the ALHAMBRA survey.

Quiescent galaxies show stronger clustering than late-type star-forming galaxies.
Their correlation function parameters at large scales are nearly compatible within the errors with fixed values around $r_0 = 5 \, h^{-1}$ Mpc and $\gamma = 2$, but the clustering length is smaller than the one calculated at low redshift by \citet{2003MNRAS.344..847M} for passive galaxies in the 2dFGRS and by \citet{2011ApJ...736...59Z} for the red galaxies in SDSS.
Instead, the values of the amplitude of the correlation function reported by \citet{Skibba2013c} at $z \sim 0.4$ for PRIMUS, by \citet{coi08a} at $z \sim 0.9$ for the DEEP2  and by \citet{Meneux2006a} agree with our results within the errors. 

For both star-forming and passive galaxies, the only discrepant measurement in Fig.~\ref{eli_seg_dre} is that corresponding to the PRIMUS samples at $z \sim 0.6$, which show values of $r_0$ significantly larger than those obtained by ALHAMBRA at similar redshifts (and also by DEEP2 at $z \sim 0.9$). 
This difference may be due to the fact that the PRIMUS survey includes the COSMOS field, which contains a large overdensity at this redshift affecting the clustering measurements (see the discussion in Section~\ref{sec:seg}).

This segregation is generally explained by the tendency of red, quiescent or early-type galaxies to form in dense environments, while blue, star-forming or late-type galaxies typically form in the field or in low mass haloes \citep{1980ApJ...236..351D,2003MNRAS.346..601G,2004ApJ...608..752B, 2005ApJ...621..673T, 2009A&A...508.1217Z, McNaught-Roberts2014a}.

\subsection{Dependence of the bias on spectral type and redshift}
\label{sec:depend-bias-spectr}

In order to disentangle the evolution of the galaxy clustering of different populations from the overall growth of structure, we study the bias $b$ of our samples based on the projected correlation function measurements.
We use a simple linear model, with a constant and scale-independent bias.
In this model, the galaxy projected correlation function is given by
\begin{equation}
\label{eq:modelbias}
w_p(r_p) = b^2 w_p^m(r_p) \, ,
\end{equation}
where $b$ is the bias, and $w^m_p(r_p)$ is the theoretical prediction for the projected correlation function of the matter distribution.
Our model for $w^m_p$ is based on $\Lambda$CDM with cosmological parameters consistent with the WMAP7 results \citep{2011ApJS..192...18K}, including a normalization of the power spectrum $\sigma_8 = 0.816$. 
The matter power spectrum at the median redshift of each sample is obtained using the \textsc{Camb} software \citep{2000ApJ...538..473L}, including the non-linear \textsc{Halofit} corrections \citep{2003MNRAS.341.1311S}.
We obtain the real-space correlation function $\xi(r)$ by a Fourier transform of the matter power spectrum and the final projected correlation function $w_p$ using eq.~(\ref{int}). 

We fit this model to our data in the range $1.0 < r_p < 10.0 \, h^{-1}$ Mpc, corresponding mainly to the two-halo term of the correlation function.
The best fit value and uncertainty of the bias is obtained by the same method as described in Section~\ref{sec:power} for the parameters of the power-law model. 
The results of these fits for each of our samples are listed in Table~\ref{fittablem}.

\begin{figure}
\begin{center}
\includegraphics[height=0.41\textheight]{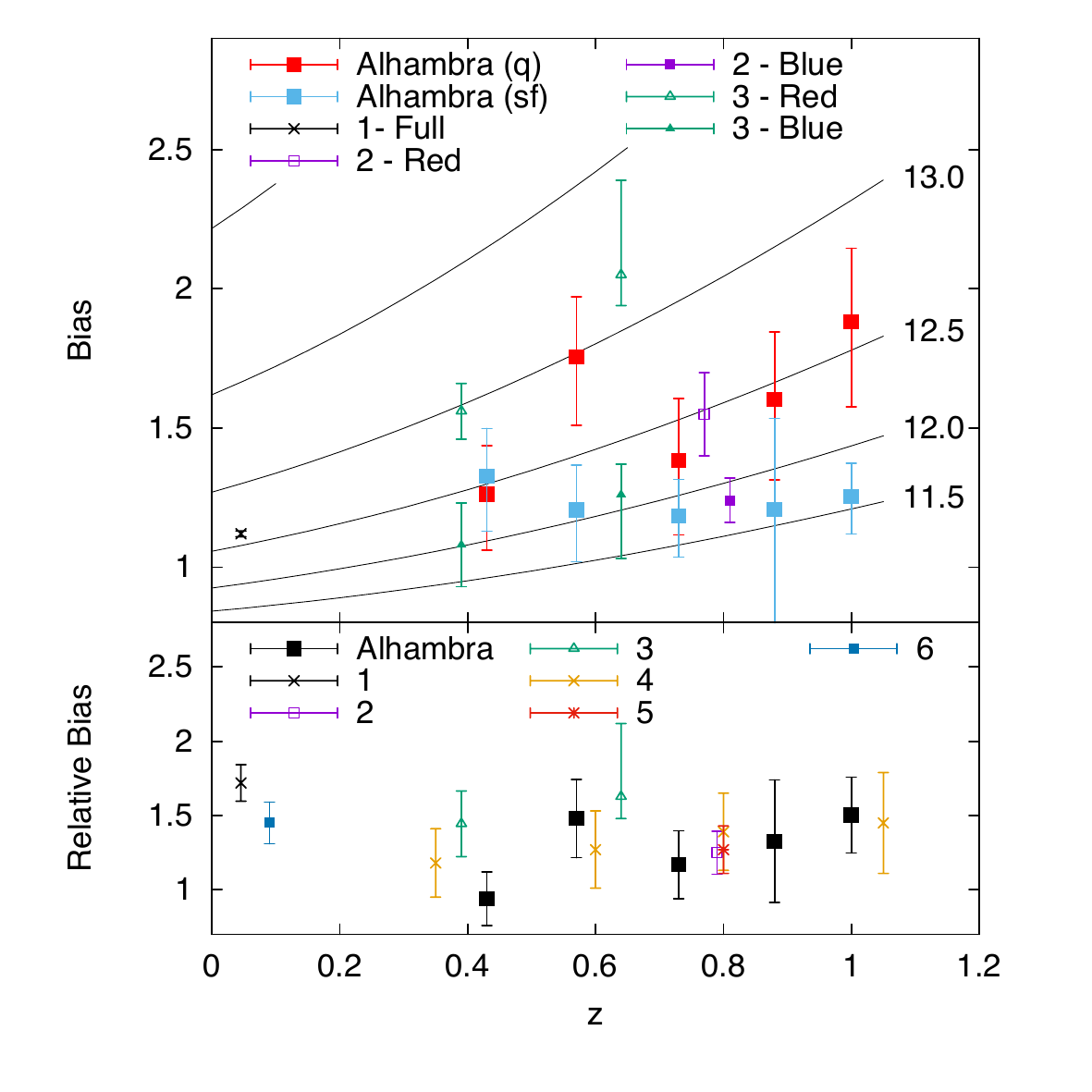}
\end{center}
\caption{\label{bias}
Galaxy bias of our quiescent (q) and star-forming (sf) galaxies as function of their median redshift.
Bias is estimated by a fit to eq.~(\ref{eq:modelbias}), as described in Sect.~\ref{sec:depend-bias-spectr}. The solid lines correspond to the bias of haloes of a fixed mass, according to the model of \citet{Tinker2005d}. These lines are labelled 
with the corresponding halo mass in terms of $\log_{10}\left[ M_{\rm h} / (h^{-1} \, {\rm M}_{\odot}) \right]$. For comparison, we plot the results obtained by 1) \cite{2011ApJ...736...59Z}, 2) \cite{coi08a}, 3) \cite{Skibba2013c}, 4) \cite{2003MNRAS.344..847M}, 5) \cite{2013Marulli}, 6) \cite{Meneux2006a} and 7) \cite{2011MNRAS.412..825D}.  Bias values from different authors have been adapted for the assumed cosmological parameters used in this paper.}
\end{figure}

We show the evolution of the bias as a function of redshift for our different populations in the top panel of Fig.~\ref{bias} (red and blue squares).  
As expected, we also see the effect of spectral segregation in this case, as the bias of early-type quiescent galaxies is consistently larger than that of late-type star-forming galaxies. 
The bias observed for the full population, not shown, is similar to that of the star-fotming galaxies.
For comparison we show, as solid lines, the bias for dark matter haloes of a fixed mass, according to the model of \citet{Tinker2005d}, and the values obtained by previous works for samples at similar redshift ranges and number densities.
The bias values in those cases were obtained by a similar method as here, and using compatible scale ranges. \footnote{In the case of \cite{2003MNRAS.344..847M}, as the bias values are not given explicitly, we derived them using their power law best fit at a scale of $5 h^{-1}$ Mpc.}.

For the star-forming galaxies, we obtain that their bias is approximately constant, with values $b \simeq 1.25$, over the range we explore. 
This explains the evolution of $r_0$ at large scales observed in Fig.~\ref{eli_seg_dre}.
If the bias is constant, the main driver for the evolution of the galaxy clustering amplitude is the growth factor, therefore $r_0$ grows with cosmic time, as observed.
Given the uncertainties, and the relatively slow evolution of the halo bias, the measured bias in this case is also consistent with the evolution of the bias of haloes with mass in the range $M_{\rm h} \simeq 10^{11.5} - 10^{12} \, h^{-1} \, \mathrm{M}_{\odot}$.
As shown in Fig.~\ref{bias}, our results for the star-forming population are fully consistent with those obtained for similar populations of blue galaxies in the the DEEP2 \citep{coi08a} and PRIMUS \citep{Skibba2013c} surveys.

The bias of quiescent galaxies shows a clear evolution, increasing with redshift, which is remarkably similar to the expected evolution of the bias for haloes of mass $M_{\rm h} \simeq 10^{12.5} \, h^{-1} \, \mathrm{M}_{\odot}$.
This clear evolution of the bias in this case compensates the clustering evolution due to the growth factor, resulting in an approximately constant value of $r_0$, as shown above in Fig.~\ref{eli_seg_dre}.
Our results for this population are consistent with the observed bias for red galaxies in the DEEP2 and PRIMUS surveys. 
We note that the bias measurement for the PRIMUS sample at $z \simeq 0.6$ may be affected by the presence of a large overdensity in the COSMOS field, as noted above.
The larger number of bins in redshift used in this ALHAMBRA analysis allows us to see more clearly this evolutionary trend.

In the bottom panel of Fig.~\ref{bias} we show the relative bias, defined as the ratio of the bias of quiescent galaxies over the bias of the star-forming ones, as function of redshift.
We also show for comparison the results from previous surveys at similar redshifts including, in addition to those shown in the top panel, the VVDS survey \citep{Meneux2006a}, and the zCOSMOS-Bright survey \citep{2011MNRAS.412..825D}. \cite{Meneux2006a} used flux-limited samples with evolving galaxy density so their absolute bias measurements are not comparable to ours. The analysis in  \cite{2011MNRAS.412..825D} only provided values of the relative bias of their samples.
At $z \sim 0$ we show the results from the 2dFGRS survey \citep{2003MNRAS.344..847M} and the SDSS \citep{2011ApJ...736...59Z}. 
We note that in the two low redshift cases, the relative bias is calculated using a slightly different method: \citet{2003MNRAS.344..847M} calculate it using the ratio of the galaxy variances $\sigma_{8, \rm gal}$ of the samples, while we have calculated the relative bias for SDSS as the ratio of the best-fit power laws of the two samples at a scale $r = 5 \, h^{-1} \, \mathrm{Mpc}$ (see, e.g. Eq.~9 of \citet{2002MNRAS.332..827N}).

We obtain values of the relative bias in the range $b_{\rm rel} \simeq 1-1.5$, consistent with all previous results at similar redshifts.
The relative bias shows a very faint evolution, slightly increasing with redshift.
However, given the errors, our results are also consistent with being constant.
A similar, faint trend is also seen for the VVDS results of \cite{Meneux2006a}.
However, if we include the $z \sim 0$ values, this evolution is broken, and the best description of the results is a constant relative bias with redshift.

Overall, our results indicate that, for samples selected by the same $B$-band luminosity and redshift, passive galaxies reside in haloes up to 10 times more massive than those hosting active galaxies.
When studying the evolution with redshift, the observed bias suggests that quiescent galaxies (following a constant number density selection) reside in haloes of constant mass, while this is not clear in the case of star-forming galaxies. 
In the latter case, there seems to be an indication that they populate slightly more massive haloes at lower redshift.
We will study the relation between galaxies and dark matter haloes using a more detailed HOD modelling in a future work.

\section{Conclusions}
\label{sec:con}

The ALHAMBRA survey allows us to perform accurate clustering calculations with different segregation criteria. Its 23-filter photometry provides reliable galaxy parameters with good completeness out to very high redshifts ($ z \sim 1.10$), opening the possibility to analyze the galaxy clustering of different galaxy populations. In AM14 the authors chose to select galaxy samples using different luminosity thresholds, while in this work we made a selection by spectral type. This selection follows the spectral classification of the ALHAMBRA photometric templates and has been proved to match remarkably well the usual selection by broad-band color. 

A rise of the correlation function at small scales is found as already noticed by \cite{coi08a}. We have been able to show that this trend holds at smaller scales and to 
characterise its redshift evolution.

Our sample allows us to measure the clustering properties of galaxy 
populations segregated by spectral type and their redshift evolution in an homogeneous 
way. At scales larger than $0.2 \, h^{-1}$ Mpc, quiescent galaxies cluster with a higher 
amplitude than star-forming ones. The difference is significant at the 2$\sigma$ level. There is also a significant hint of evolution (2$\sigma$) in the clustering amplitude of active galaxies, while the clustering of the passive ones remains 
constant. These results are compatible with previous works in the literature, but 
in the present work we have increased the redshift resolution. 

Regarding the small scales ($r_p < 0.2 \, h^{-1}$ Mpc) we find almost no change in the 
correlation function compared with the large scales in the quiescent population. On 
the other hand, the star-forming galaxies show a clear variation in the slope 
between small and large scales, which is possibly decreasing towards low 
redshifts. 

Our measurements of the bias value for the different populations show strong 
segregation between them. The bias of the quiescent population clearly evolves 
with redshift following the expected behaviour for haloes of approximate mass 
$10^{12.5} \, h^{-1} M_{\astrosun}$.  The star-forming population bias remains basically constant over our observed redshift range, but it can still be compatible with the theoretical 
evolution of lower mass haloes ($M_h \sim 10^{11.5-12}  \, h^{-1} M_{\astrosun}$). As a consequence, the relative bias hints at a slow evolution, which would not be completely consistent with observations at $z \sim 0$.

\acknowledgements
We would like to thank an anonymous referee for his/her comments that have improved the quality and readability of this paper. This work is based on observations collected at the German–Spanish Astronomical Center, Calar Alto, jointly operated by the Max-Planck-Institut f\"ur Astronomie (MPIA) and the Instituto de Astrof\'{\i}sica de Andaluc\'{\i}a (CSIC). This work was mainly supported by the Spanish Ministry for Economy and Competitiveness and FEDER funds through grants AYA2010-22111-C03-02 and AYA2013-48623-C2-2, and by the Generalitat Valenciana through project PrometeoII 2014/060.
We also acknowledge support from the Spanish Ministry for Economy and Competitiveness and FEDER funds through grants  AYA2012-39620, AYA2013-40611-P, AYA2013-42227-P, AYA2013-43188-P, AYA2013-48623-C2-1, ESP2013-48274, AYA2014-58861-C3-1,  Junta de Andaluc\'{\i}a grants TIC114, JA2828, P10-FQM-6444, and Generalitat de Catalunya project SGR-1398. 
BA acknowledges received funding from the European Union's Horizon 2020 research and innovation programme under the Marie Sklodowska-Curie grant agreement No 656354.

\bibliography{ref}

\begin{thebibliography}{}
\expandafter\ifx\csname natexlab\endcsname\relax\def\natexlab#1{#1}\fi

\bibitem[{{Abbas} \& {Sheth}(2006)}]{2006MNRAS.372.1749A}
{Abbas}, U., \& {Sheth}, R.~K. 2006, \mnras, 372, 1749

\bibitem[{{Arnalte-Mur} {et~al.}(2009){Arnalte-Mur}, {Fern{\'a}ndez-Soto},
  {Mart{\'{\i}}nez}, {Saar}, {Hein{\"a}m{\"a}ki}, \&
  {Suhhonenko}}]{2009MNRAS.394.1631A}
{Arnalte-Mur}, P., {Fern{\'a}ndez-Soto}, A., {Mart{\'{\i}}nez}, V.~J., {et~al.}
  2009, \mnras, 394, 1631

\bibitem[{{Arnalte-Mur} {et~al.}(2014){Arnalte-Mur}, {Mart{\'{\i}}nez},
  {Norberg}, {Fern{\'a}ndez-Soto}, {Ascaso}, {Merson}, {Aguerri}, {Castander},
  {Hurtado-Gil}, {L{\'o}pez-Sanjuan}, {Molino}, {Montero-Dorta}, {Stefanon},
  {Alfaro}, {Aparicio-Villegas}, {Ben{\'{\i}}tez}, {Broadhurst},
  {Cabrera-Ca{\~n}o}, {Cepa}, {Cervi{\~n}o}, {Crist{\'o}bal-Hornillos}, {del
  Olmo}, {Gonz{\'a}lez Delgado}, {Husillos}, {Infante}, {M{\'a}rquez},
  {Masegosa}, {Moles}, {Perea}, {Povi{\'c}}, {Prada}, \&
  {Quintana}}]{2014MNRAS.441.1783A}
{Arnalte-Mur}, P., {Mart{\'{\i}}nez}, V.~J., {Norberg}, P., {et~al.} 2014,
  \mnras, 441, 1783

\bibitem[{{Ascaso} {et~al.}(2015){Ascaso}, {Ben{\'{\i}}tez},
  {Fern{\'a}ndez-Soto}, {Arnalte-Mur}, {L{\'o}pez-Sanjuan}, {Molino},
  {Schoenell}, {Jim{\'e}nez-Teja}, {Merson}, {Huertas-Company},
  {D{\'{\i}}az-Garc{\'{\i}}a}, {Mart{\'{\i}}nez}, {Cenarro}, {Dupke},
  {M{\'a}rquez}, {Masegosa}, {Nieves-Seoane}, {Povi{\'c}}, {Varela},
  {Viironen}, {Aguerri}, {Olmo}, {Moles}, {Perea}, {Alfaro},
  {Aparicio-Villegas}, {Broadhurst}, {Cabrera-Ca{\~n}o}, {Castander}, {Cepa},
  {Cervi{\~n}o}, {Delgado}, {Crist{\'o}bal-Hornillos}, {Hurtado-Gil},
  {Husillos}, {Infante}, {Prada}, \& {Quintana}}]{Ascaso2015a}
{Ascaso}, B., {Ben{\'{\i}}tez}, N., {Fern{\'a}ndez-Soto}, A., {et~al.} 2015,
  \mnras, 452, 549

\bibitem[{{Bell} {et~al.}(2004){Bell}, {Wolf}, {Meisenheimer}, {Rix}, {Borch},
  {Dye}, {Kleinheinrich}, {Wisotzki}, \& {McIntosh}}]{2004ApJ...608..752B}
{Bell}, E.~F., {Wolf}, C., {Meisenheimer}, K., {et~al.} 2004, \apj, 608, 752

\bibitem[{{Ben{\'{\i}}tez}(2000)}]{2000ApJ...536..571B}
{Ben{\'{\i}}tez}, N. 2000, \apj, 536, 571

\bibitem[{{Coil} {et~al.}(2006){Coil}, {Newman}, {Cooper}, {Davis}, {Faber},
  {Koo}, \& {Willmer}}]{2006ApJ...644..671C}
{Coil}, A.~L., {Newman}, J.~A., {Cooper}, M.~C., {et~al.} 2006, \apj, 644, 671

\bibitem[{{Coil} {et~al.}(2008){Coil}, {Newman}, {Croton}, {Cooper}, {Davis},
  {Faber}, {Gerke}, {Koo}, {Padmanabhan}, {Wechsler}, \& {Weiner}}]{coi08a}
{Coil}, A.~L., {Newman}, J.~A., {Croton}, D., {et~al.} 2008, \apj, 672, 153

\bibitem[{{Cucciati} {et~al.}(2006){Cucciati}, {Iovino}, {Marinoni}, {Ilbert},
  {Bardelli}, {Franzetti}, {Le F{\`e}vre}, {Pollo}, {Zamorani}, {Cappi},
  {Guzzo}, {McCracken}, {Meneux}, {Scaramella}, {Scodeggio}, {Tresse}, {Zucca},
  {Bottini}, {Garilli}, {Le Brun}, {Maccagni}, {Picat}, {Vettolani},
  {Zanichelli}, {Adami}, {Arnaboldi}, {Arnouts}, {Bolzonella}, {Charlot},
  {Ciliegi}, {Contini}, {Foucaud}, {Gavignaud}, {Marano}, {Mazure}, {Merighi},
  {Paltani}, {Pell{\`o}}, {Pozzetti}, {Radovich}, {Bondi}, {Bongiorno},
  {Busarello}, {de la Torre}, {Gregorini}, {Lamareille}, {Mathez}, {Mellier},
  {Merluzzi}, {Ripepi}, {Rizzo}, {Temporin}, \&
  {Vergani}}]{2006A&A...458...39C}
{Cucciati}, O., {Iovino}, A., {Marinoni}, C., {et~al.} 2006, \aap, 458, 39

\bibitem[{{Davis} \& {Geller}(1976)}]{1976ApJ...208...13D}
{Davis}, M., \& {Geller}, M.~J. 1976, \apj, 208, 13

\bibitem[{{Davis} {et~al.}(1988){Davis}, {Meiksin}, {Strauss}, {da Costa}, \&
  {Yahil}}]{1988ApJ...333L...9D}
{Davis}, M., {Meiksin}, A., {Strauss}, M.~A., {da Costa}, L.~N., \& {Yahil}, A.
  1988, \apjl, 333, L9

\bibitem[{{Davis} \& {Peebles}(1983)}]{1983ApJ...267..465D}
{Davis}, M., \& {Peebles}, P.~J.~E. 1983, \apj, 267, 465

\bibitem[{{de la Torre} {et~al.}(2010){de la Torre}, {Guzzo}, {Kova{\v c}},
  {Porciani}, {Abbas}, {Meneux}, {Carollo}, {Contini}, {Kneib}, {Le F{\`e}vre},
  {Lilly}, {Mainieri}, {Renzini}, {Sanders}, {Scodeggio}, {Scoville},
  {Zamorani}, {Bardelli}, {Bolzonella}, {Bongiorno}, {Caputi}, {Coppa},
  {Cucciati}, {de Ravel}, {Franzetti}, {Garilli}, {Iovino}, {Kampczyk},
  {Knobel}, {Koekemoer}, {Lamareille}, {Le Borgne}, {Le Brun}, {Maier},
  {Mignoli}, {Pell{\'o}}, {Peng}, {Perez-Montero}, {Ricciardelli}, {Silverman},
  {Tanaka}, {Tasca}, {Tresse}, {Vergani}, {Welikala}, {Zucca}, {Bottini},
  {Cappi}, {Cassata}, {Cimatti}, {Fumana}, {Ilbert}, {Leauthaud}, {Maccagni},
  {Marinoni}, {McCracken}, {Memeo}, {Nair}, {Oesch}, {Pozzetti}, {Presotto}, \&
  {Scaramella}}]{tor10a}
{de la Torre}, S., {Guzzo}, L., {Kova{\v c}}, K., {et~al.} 2010, \mnras, 409,
  867

\bibitem[{de~la Torre {et~al.}(2011)de~la Torre, {Le F{\`e}vre}, {Porciani},
  {Guzzo}, {Meneux}, {Abbas}, {Tasca}, {Carollo}, {Contini}, {Kneib}, {Lilly},
  {Mainieri}, {Renzini}, {Scodeggio}, {Zamorani}, {Bardelli}, {Bolzonella},
  {Bongiorno}, {Caputi}, {Coppa}, {Cucciati}, {de Ravel}, {Franzetti},
  {Garilli}, {Halliday}, {Iovino}, {Kampczyk}, {Knobel}, {Koekemoer}, {Kova{\v
  c}}, {Lamareille}, {Le Borgne}, {Le Brun}, {Maier}, {Mignoli}, {Pell{\'o}},
  {Peng}, {Perez-Montero}, {Ricciardelli}, {Silverman}, {Tanaka}, {Tresse},
  {Vergani}, {Zucca}, {Bottini}, {Cappi}, {Cassata}, {Cimatti}, {Leauthaud},
  {Maccagni}, {Marinoni}, {McCracken}, {Memeo}, {Oesch}, {Pozzetti}, \&
  {Scaramella}}]{tor09a}
de~la Torre, S., {Le F{\`e}vre}, O., {Porciani}, C., {et~al.} 2011, \mnras,
  412, 825

\bibitem[{{de la Torre} {et~al.}(2011){de la Torre}, {Le F{\`e}vre},
  {Porciani}, {Guzzo}, {Meneux}, {Abbas}, {Tasca}, {Carollo}, {Contini},
  {Kneib}, {Lilly}, {Mainieri}, {Renzini}, {Scodeggio}, {Zamorani}, {Bardelli},
  {Bolzonella}, {Bongiorno}, {Caputi}, {Coppa}, {Cucciati}, {de Ravel},
  {Franzetti}, {Garilli}, {Halliday}, {Iovino}, {Kampczyk}, {Knobel},
  {Koekemoer}, {Kova{\v c}}, {Lamareille}, {Le Borgne}, {Le Brun}, {Maier},
  {Mignoli}, {Pell{\'o}}, {Peng}, {Perez-Montero}, {Ricciardelli}, {Silverman},
  {Tanaka}, {Tresse}, {Vergani}, {Zucca}, {Bottini}, {Cappi}, {Cassata},
  {Cimatti}, {Leauthaud}, {Maccagni}, {Marinoni}, {McCracken}, {Memeo},
  {Oesch}, {Pozzetti}, \& {Scaramella}}]{2011MNRAS.412..825D}
{de la Torre}, S., {Le F{\`e}vre}, O., {Porciani}, C., {et~al.} 2011, \mnras,
  412, 825

\bibitem[{{Dom\'{\i}nguez-Tenreiro} \&
  {Mart\'{\i}nez}(1989)}]{1989ApJ...339L...9D}
{Dom\'{\i}nguez-Tenreiro}, R., \& {Mart\'{\i}nez}, V.~J. 1989, \apjl, 339, L9

\bibitem[{{Dressler}(1980)}]{1980ApJ...236..351D}
{Dressler}, A. 1980, \apj, 236, 351

\bibitem[{{Einasto}(1991)}]{1991MNRAS.252..261E}
{Einasto}, M. 1991, \mnras, 252, 261

\bibitem[{{Giovanelli} {et~al.}(1986){Giovanelli}, {Haynes}, \&
  {Chincarini}}]{1986ApJ...300...77G}
{Giovanelli}, R., {Haynes}, M.~P., \& {Chincarini}, G.~L. 1986, \apj, 300, 77

\bibitem[{{Goto} {et~al.}(2003){Goto}, {Yamauchi}, {Fujita}, {Okamura},
  {Sekiguchi}, {Smail}, {Bernardi}, \& {Gomez}}]{2003MNRAS.346..601G}
{Goto}, T., {Yamauchi}, C., {Fujita}, Y., {et~al.} 2003, \mnras, 346, 601

\bibitem[{{Guo} {et~al.}(2012){Guo}, {Zehavi}, \&
  {Zheng}}]{2012ApJ...756..127G}
{Guo}, H., {Zehavi}, I., \& {Zheng}, Z. 2012, \apj, 756, 127

\bibitem[{{Guo} {et~al.}(2013){Guo}, {Zehavi}, {Zheng}, {Weinberg}, {Berlind},
  {Blanton}, {Chen}, {Eisenstein}, {Ho}, {Kazin}, {Manera}, {Maraston},
  {McBride}, {Nuza}, {Padmanabhan}, {Parejko}, {Percival}, {Ross}, {Ross},
  {Samushia}, {S{\'a}nchez}, {Schlegel}, {Schneider}, {Skibba}, {Swanson},
  {Tinker}, {Tojeiro}, {Wake}, {White}, {Bahcall}, {Bizyaev}, {Brewington},
  {Bundy}, {da Costa}, {Ebelke}, {Malanushenko}, {Malanushenko}, {Oravetz},
  {Rossi}, {Simmons}, {Snedden}, {Streblyanska}, \&
  {Thomas}}]{2013ApJ...767..122G}
{Guo}, H., {Zehavi}, I., {Zheng}, Z., {et~al.} 2013, \apj, 767, 122

\bibitem[{{Guzzo} {et~al.}(1997){Guzzo}, {Strauss}, {Fisher}, {Giovanelli}, \&
  {Haynes}}]{1997ApJ...489...37G}
{Guzzo}, L., {Strauss}, M.~A., {Fisher}, K.~B., {Giovanelli}, R., \& {Haynes},
  M.~P. 1997, \apj, 489, 37

\bibitem[{{Guzzo} \& {The Vipers Team}(2013)}]{2013Msngr.151...41G}
{Guzzo}, L., \& {The Vipers Team}. 2013, The Messenger, 151, 41

\bibitem[{{Guzzo} {et~al.}(2007){Guzzo}, {Cassata}, {Finoguenov}, {Massey},
  {Scoville}, {Capak}, {Ellis}, {Mobasher}, {Taniguchi}, {Thompson}, {Ajiki},
  {Aussel}, {B{\"o}hringer}, {Brusa}, {Calzetti}, {Comastri}, {Franceschini},
  {Hasinger}, {Kasliwal}, {Kitzbichler}, {Kneib}, {Koekemoer}, {Leauthaud},
  {McCracken}, {Murayama}, {Nagao}, {Rhodes}, {Sanders}, {Sasaki}, {Shioya},
  {Tasca}, \& {Taylor}}]{Guzzo2007a}
{Guzzo}, L., {Cassata}, P., {Finoguenov}, A., {et~al.} 2007, \apjs, 172, 254

\bibitem[{{Hamilton}(1988)}]{1988ApJ...331L..59H}
{Hamilton}, A.~J.~S. 1988, \apjl, 331, L59

\bibitem[{{Hamilton} \& {Tegmark}(2004)}]{ham04a}
{Hamilton}, A.~J.~S., \& {Tegmark}, M. 2004, \mnras, 349, 115

\bibitem[{{Hartley} {et~al.}(2010){Hartley}, {Almaini}, {Cirasuolo}, {Foucaud},
  {Simpson}, {Conselice}, {Smail}, {McLure}, {Dunlop}, {Chuter}, {Maddox},
  {Lane}, \& {Bradshaw}}]{har10a}
{Hartley}, W.~G., {Almaini}, O., {Cirasuolo}, M., {et~al.} 2010, \mnras, 407,
  1212

\bibitem[{{Komatsu} {et~al.}(2011){Komatsu}, {Smith}, {Dunkley}, {Bennett},
  {Gold}, {Hinshaw}, {Jarosik}, {Larson}, {Nolta}, {Page}, {Spergel},
  {Halpern}, {Hill}, {Kogut}, {Limon}, {Meyer}, {Odegard}, {Tucker}, {Weiland},
  {Wollack}, \& {Wright}}]{2011ApJS..192...18K}
{Komatsu}, E., {Smith}, K.~M., {Dunkley}, J., {et~al.} 2011, \apjs, 192, 18

\bibitem[{{Landy} \& {Szalay}(1993)}]{1993ApJ...412...64L}
{Landy}, S.~D., \& {Szalay}, A.~S. 1993, \apj, 412, 64

\bibitem[{{Lewis} {et~al.}(2000){Lewis}, {Challinor}, \&
  {Lasenby}}]{2000ApJ...538..473L}
{Lewis}, A., {Challinor}, A., \& {Lasenby}, A. 2000, \apj, 538, 473

\bibitem[{{Li} {et~al.}(2006){Li}, {Kauffmann}, {Jing}, {White}, {B{\"o}rner},
  \& {Cheng}}]{2006MNRAS.368...21L}
{Li}, C., {Kauffmann}, G., {Jing}, Y.~P., {et~al.} 2006, \mnras, 368, 21

\bibitem[{{Loveday} {et~al.}(1995){Loveday}, {Maddox}, {Efstathiou}, \&
  {Peterson}}]{1995ApJ...442..457L}
{Loveday}, J., {Maddox}, S.~J., {Efstathiou}, G., \& {Peterson}, B.~A. 1995,
  \apj, 442, 457

\bibitem[{{Madgwick} {et~al.}(2003){Madgwick}, {Hawkins}, {Lahav}, {Maddox},
  {Norberg}, {Peacock}, {Baldry}, {Baugh}, {Bland-Hawthorn}, {Bridges},
  {Cannon}, {Cole}, {Colless}, {Collins}, {Couch}, {Dalton}, {De Propris},
  {Driver}, {Efstathiou}, {Ellis}, {Frenk}, {Glazebrook}, {Jackson}, {Lewis},
  {Lumsden}, {Peterson}, {Sutherland}, \& {Taylor}}]{2003MNRAS.344..847M}
{Madgwick}, D.~S., {Hawkins}, E., {Lahav}, O., {et~al.} 2003, \mnras, 344, 847

\bibitem[{{Mart{\'{\i}}nez} {et~al.}(2010){Mart{\'{\i}}nez}, {Arnalte-Mur}, \&
  {Stoyan}}]{2010A&A...513A..22M}
{Mart{\'{\i}}nez}, V.~J., {Arnalte-Mur}, P., \& {Stoyan}, D. 2010, \aap, 513,
  A22

\bibitem[{{Mart{\'{\i}}nez} \& {Saar}(2002)}]{2002sgd..book.....M}
{Mart{\'{\i}}nez}, V.~J., \& {Saar}, E. 2002, {Statistics of the Galaxy
  Distribution} (Chapman \&amp)

\bibitem[{{Marulli} {et~al.}(2013){Marulli}, {Bolzonella}, {Branchini},
  {Davidzon}, {de la Torre}, {Granett}, {Guzzo}, {Iovino}, {Moscardini},
  {Pollo}, {Abbas}, {Adami}, {Arnouts}, {Bel}, {Bottini}, {Cappi}, {Coupon},
  {Cucciati}, {De Lucia}, {Fritz}, {Franzetti}, {Fumana}, {Garilli}, {Ilbert},
  {Krywult}, {Le Brun}, {Le F{\`e}vre}, {Maccagni}, {Ma{\l}ek}, {McCracken},
  {Paioro}, {Polletta}, {Schlagenhaufer}, {Scodeggio}, {Tasca}, {Tojeiro},
  {Vergani}, {Zanichelli}, {Burden}, {Di Porto}, {Marchetti}, {Marinoni},
  {Mellier}, {Nichol}, {Peacock}, {Percival}, {Phleps}, {Wolk}, \&
  {Zamorani}}]{2013Marulli}
{Marulli}, F., {Bolzonella}, M., {Branchini}, E., {et~al.} 2013, \aap, 557, A17

\bibitem[{{Matthews} \& {Newman}(2012)}]{2012ApJ...745..180M}
{Matthews}, D.~J., \& {Newman}, J.~A. 2012, \apj, 745, 180

\bibitem[{{McCracken} {et~al.}(2007){McCracken}, {Peacock}, {Guzzo}, {Capak},
  {Porciani}, {Scoville}, {Aussel}, {Finoguenov}, {James}, {Kitzbichler},
  {Koekemoer}, {Leauthaud}, {Le F{\`e}vre}, {Massey}, {Mellier}, {Mobasher},
  {Norberg}, {Rhodes}, {Sanders}, {Sasaki}, {Taniguchi}, {Thompson}, {White},
  \& {El-Zant}}]{McCracken2007a}
{McCracken}, H.~J., {Peacock}, J.~A., {Guzzo}, L., {et~al.} 2007, \apjs, 172,
  314

\bibitem[{{McCracken} {et~al.}(2015){McCracken}, {Wolk}, {Colombi},
  {Kilbinger}, {Ilbert}, {Peirani}, {Coupon}, {Dunlop}, {Milvang-Jensen},
  {Caputi}, {Aussel}, {B{\'e}thermin}, \& {Le F{\`e}vre}}]{McCracken2015c}
{McCracken}, H.~J., {Wolk}, M., {Colombi}, S., {et~al.} 2015, \mnras, 449, 901

\bibitem[{{McNaught-Roberts} {et~al.}(2014){McNaught-Roberts}, {Norberg},
  {Baugh}, {Lacey}, {Loveday}, {Peacock}, {Baldry}, {Bland-Hawthorn}, {Brough},
  {Driver}, {Robotham}, \& {V{\'a}zquez-Mata}}]{McNaught-Roberts2014a}
{McNaught-Roberts}, T., {Norberg}, P., {Baugh}, C., {et~al.} 2014, \mnras, 445,
  2125

\bibitem[{{Meneux} {et~al.}(2006){Meneux}, {Le F{\`e}vre}, {Guzzo}, {Pollo},
  {Cappi}, {Ilbert}, {Iovino}, {Marinoni}, {McCracken}, {Bottini}, {Garilli},
  {Le Brun}, {Maccagni}, {Picat}, {Scaramella}, {Scodeggio}, {Tresse},
  {Vettolani}, {Zanichelli}, {Adami}, {Arnouts}, {Arnaboldi}, {Bardelli},
  {Bolzonella}, {Charlot}, {Ciliegi}, {Contini}, {Foucaud}, {Franzetti},
  {Gavignaud}, {Marano}, {Mazure}, {Merighi}, {Paltani}, {Pell{\`o}},
  {Pozzetti}, {Radovich}, {Zamorani}, {Zucca}, {Bondi}, {Bongiorno},
  {Busarello}, {Cucciati}, {Gregorini}, {Lamareille}, {Mathez}, {Mellier},
  {Merluzzi}, {Ripepi}, \& {Rizzo}}]{Meneux2006a}
{Meneux}, B., {Le F{\`e}vre}, O., {Guzzo}, L., {et~al.} 2006, \aap, 452, 387

\bibitem[{{Moles} {et~al.}(2008){Moles}, {Ben{\'{\i}}tez}, {Aguerri}, {Alfaro},
  {Broadhurst}, {Cabrera-Ca{\~n}o}, {Castander}, {Cepa}, {Cervi{\~n}o},
  {Crist{\'o}bal-Hornillos}, {Fern{\'a}ndez-Soto}, {Gonz{\'a}lez Delgado},
  {Infante}, {M{\'a}rquez}, {Mart{\'{\i}}nez}, {Masegosa}, {del Olmo}, {Perea},
  {Prada}, {Quintana}, \& {S{\'a}nchez}}]{mol08a}
{Moles}, M., {Ben{\'{\i}}tez}, N., {Aguerri}, J.~A.~L., {et~al.} 2008, \aj,
  136, 1325

\bibitem[{{Molino} {et~al.}(2014){Molino}, {Ben{\'{\i}}tez}, {Moles},
  {Fern{\'a}ndez-Soto}, {Crist{\'o}bal-Hornillos}, {Ascaso},
  {Jim{\'e}nez-Teja}, {Schoenell}, {Arnalte-Mur}, {Povi{\'c}}, {Coe},
  {L{\'o}pez-Sanjuan}, {D{\'{\i}}az-Garc{\'{\i}}a}, {Varela}, {Stefanon},
  {Cenarro}, {Matute}, {Masegosa}, {M{\'a}rquez}, {Perea}, {Del Olmo},
  {Husillos}, {Alfaro}, {Aparicio-Villegas}, {Cervi{\~n}o}, {Huertas-Company},
  {Aguerri}, {Broadhurst}, {Cabrera-Ca{\~n}o}, {Cepa}, {Gonz{\'a}lez},
  {Infante}, {Mart{\'{\i}}nez}, {Prada}, \& {Quintana}}]{2013arXiv1306.4968M}
{Molino}, A., {Ben{\'{\i}}tez}, N., {Moles}, M., {et~al.} 2014, \mnras, 441,
  2891

\bibitem[{{Norberg} {et~al.}(2009){Norberg}, {Baugh}, {Gazta{\~n}aga}, \&
  {Croton}}]{nor08a}
{Norberg}, P., {Baugh}, C.~M., {Gazta{\~n}aga}, E., \& {Croton}, D.~J. 2009,
  \mnras, 396, 19

\bibitem[{{Norberg} {et~al.}(2002){Norberg}, {Baugh}, {Hawkins}, {Maddox},
  {Madgwick}, {Lahav}, {Cole}, {Frenk}, {Baldry}, {Bland-Hawthorn}, {Bridges},
  {Cannon}, {Colless}, {Collins}, {Couch}, {Dalton}, {De Propris}, {Driver},
  {Efstathiou}, {Ellis}, {Glazebrook}, {Jackson}, {Lewis}, {Lumsden},
  {Peacock}, {Peterson}, {Sutherland}, \& {Taylor}}]{2002MNRAS.332..827N}
{Norberg}, P., {Baugh}, C.~M., {Hawkins}, E., {et~al.} 2002, \mnras, 332, 827

\bibitem[{{Peebles}(1980)}]{1980lssu.book.....P}
{Peebles}, P.~J.~E. 1980, {The large-scale structure of the universe}
  (Princeton university press)

\bibitem[{{Phleps} {et~al.}(2006){Phleps}, {Peacock}, {Meisenheimer}, \&
  {Wolf}}]{2006A&A...457..145P}
{Phleps}, S., {Peacock}, J.~A., {Meisenheimer}, K., \& {Wolf}, C. 2006, \aap,
  457, 145

\bibitem[{{Povi{\'c}} {et~al.}(2013){Povi{\'c}}, {Huertas-Company}, {Aguerri},
  {M{\'a}rquez}, {Masegosa}, {Husillos}, {Molino}, {Crist{\'o}bal-Hornillos},
  {Perea}, {Ben{\'{\i}}tez}, {Olmo}, {Fern{\'a}ndez-Soto}, {Jim{\'e}nez-Teja},
  {Moles}, {Alfaro}, {Aparicio-Villegas}, {Ascaso}, {Broadhurst},
  {Cabrera-Ca{\~n}o}, {Castander}, {Cepa}, {Fernandez Lorenzo}, {Cervi{\~n}o},
  {Delgado}, {Infante}, {L{\'o}pez-Sanjuan}, {Mart{\'{\i}}nez}, {Matute},
  {Oteo}, {P{\'e}rez-Garc{\'{\i}}a}, {Prada}, \&
  {Quintana}}]{2013MNRAS.435.3444P}
{Povi{\'c}}, M., {Huertas-Company}, M., {Aguerri}, J.~A.~L., {et~al.} 2013,
  \mnras, 435, 3444

\bibitem[{{Roche} {et~al.}(1999){Roche}, {Eales}, {Hippelein}, \&
  {Willott}}]{Roche1999o}
{Roche}, N., {Eales}, S.~A., {Hippelein}, H., \& {Willott}, C.~J. 1999, \mnras,
  306, 538

\bibitem[{{Scoville} {et~al.}(2007{\natexlab{a}}){Scoville}, {Aussel},
  {Benson}, {Blain}, {Calzetti}, {Capak}, {Ellis}, {El-Zant}, {Finoguenov},
  {Giavalisco}, {Guzzo}, {Hasinger}, {Koda}, {Le F{\`e}vre}, {Massey},
  {McCracken}, {Mobasher}, {Renzini}, {Rhodes}, {Salvato}, {Sanders}, {Sasaki},
  {Schinnerer}, {Sheth}, {Shopbell}, {Taniguchi}, {Taylor}, \&
  {Thompson}}]{Scoville2007d}
{Scoville}, N., {Aussel}, H., {Benson}, A., {et~al.} 2007{\natexlab{a}}, \apjs,
  172, 150

\bibitem[{{Scoville} {et~al.}(2007{\natexlab{b}}){Scoville}, {Aussel}, {Brusa},
  {Capak}, {Carollo}, {Elvis}, {Giavalisco}, {Guzzo}, {Hasinger}, {Impey},
  {Kneib}, {LeFevre}, {Lilly}, {Mobasher}, {Renzini}, {Rich}, {Sanders},
  {Schinnerer}, {Schminovich}, {Shopbell}, {Taniguchi}, \&
  {Tyson}}]{Scoville2007h}
{Scoville}, N., {Aussel}, H., {Brusa}, M., {et~al.} 2007{\natexlab{b}}, \apjs,
  172, 1

\bibitem[{{Seljak}(2000)}]{2000MNRAS.318..203S}
{Seljak}, U. 2000, \mnras, 318, 203

\bibitem[{{Skibba} {et~al.}(2014){Skibba}, {Smith}, {Coil}, {Moustakas},
  {Aird}, {Blanton}, {Bray}, {Cool}, {Eisenstein}, {Mendez}, {Wong}, \&
  {Zhu}}]{Skibba2013c}
{Skibba}, R.~A., {Smith}, M.~S.~M., {Coil}, A.~L., {et~al.} 2014, \apj, 784,
  128

\bibitem[{{Smith} {et~al.}(2003){Smith}, {Peacock}, {Jenkins}, {White},
  {Frenk}, {Pearce}, {Thomas}, {Efstathiou}, \&
  {Couchman}}]{2003MNRAS.341.1311S}
{Smith}, R.~E., {Peacock}, J.~A., {Jenkins}, A., {et~al.} 2003, \mnras, 341,
  1311

\bibitem[{{Stefanon}(2011)}]{2011Stefanon}
{Stefanon}, M. 2011, PhD Thesis, Universitat de Valencia, 1, 1

\bibitem[{{Swanson} {et~al.}(2008){Swanson}, {Tegmark}, {Hamilton}, \&
  {Hill}}]{swa08a}
{Swanson}, M.~E.~C., {Tegmark}, M., {Hamilton}, A.~J.~S., \& {Hill}, J.~C.
  2008, \mnras, 387, 1391

\bibitem[{{Thomas} {et~al.}(2005){Thomas}, {Maraston}, {Bender}, \& {Mendes de
  Oliveira}}]{2005ApJ...621..673T}
{Thomas}, D., {Maraston}, C., {Bender}, R., \& {Mendes de Oliveira}, C. 2005,
  \apj, 621, 673

\bibitem[{{Tinker} {et~al.}(2005){Tinker}, {Weinberg}, {Zheng}, \&
  {Zehavi}}]{Tinker2005d}
{Tinker}, J.~L., {Weinberg}, D.~H., {Zheng}, Z., \& {Zehavi}, I. 2005, \apj,
  631, 41

\bibitem[{{Zehavi} {et~al.}(2004){Zehavi}, {Weinberg}, {Zheng}, {Berlind},
  {Frieman}, {Scoccimarro}, {Sheth}, {Blanton}, {Tegmark}, {Mo}, {Bahcall},
  {Brinkmann}, {Burles}, {Csabai}, {Fukugita}, {Gunn}, {Lamb}, {Loveday},
  {Lupton}, {Meiksin}, {Munn}, {Nichol}, {Schlegel}, {Schneider}, {SubbaRao},
  {Szalay}, {Uomoto}, {York}, \& {SDSS Collaboration}}]{2004ApJ...608...16Z}
{Zehavi}, I., {Weinberg}, D.~H., {Zheng}, Z., {et~al.} 2004, \apj, 608, 16

\bibitem[{{Zehavi} {et~al.}(2011){Zehavi}, {Zheng}, {Weinberg}, {Blanton},
  {Bahcall}, {Berlind}, {Brinkmann}, {Frieman}, {Gunn}, {Lupton}, {Nichol},
  {Percival}, {Schneider}, {Skibba}, {Strauss}, {Tegmark}, \&
  {York}}]{2011ApJ...736...59Z}
{Zehavi}, I., {Zheng}, Z., {Weinberg}, D.~H., {et~al.} 2011, \apj, 736, 59

\bibitem[{{Zucca} {et~al.}(2009){Zucca}, {Bardelli}, {Bolzonella}, {Zamorani},
  {Ilbert}, {Pozzetti}, {Mignoli}, {Kova{\v c}}, {Lilly}, {Tresse}, {Tasca},
  {Cassata}, {Halliday}, {Vergani}, {Caputi}, {Carollo}, {Contini}, {Kneib},
  {Le F{\`e}vre}, {Mainieri}, {Renzini}, {Scodeggio}, {Bongiorno}, {Coppa},
  {Cucciati}, {de La Torre}, {de Ravel}, {Franzetti}, {Garilli}, {Iovino},
  {Kampczyk}, {Knobel}, {Lamareille}, {Le Borgne}, {Le Brun}, {Maier},
  {Pell{\`o}}, {Peng}, {Perez-Montero}, {Ricciardelli}, {Silverman}, {Tanaka},
  {Abbas}, {Bottini}, {Cappi}, {Cimatti}, {Guzzo}, {Koekemoer}, {Leauthaud},
  {Maccagni}, {Marinoni}, {McCracken}, {Memeo}, {Meneux}, {Moresco}, {Oesch},
  {Porciani}, {Scaramella}, {Arnouts}, {Aussel}, {Capak}, {Kartaltepe},
  {Salvato}, {Sanders}, {Scoville}, {Taniguchi}, \&
  {Thompson}}]{2009A&A...508.1217Z}
{Zucca}, E., {Bardelli}, S., {Bolzonella}, M., {et~al.} 2009, \aap, 508, 1217

\end{thebibliography}

\section*{Author Affiliations}
\small{
\noindent
$^1$ Observatori Astron\`omic, Universitat de Val\`encia, C/ Catedr\`atic Jos\'e Beltr\'an 2, E-46980, Paterna, Spain\\
$^2$ Instituto de F\'isica de Cantabria (CSIC-UC), E-39005 Santander, Spain\\
$^3$ Departament d'Astronomia i Astrof\'isica, Universitat de Val\`encia, E-46100, Burjassot, Spain\\
$^4$ Unidad Asociada Observatorio Astron\'omico (IFCA-UV), E-46980, Paterna, Spain\\
$^5$ Leiden Observatory, Leiden University, P.O. Box 9513, 2300 RA Leiden, The Netherlands\\
$^6$ GEPI, Observatoire de Paris, CNRS, Universit\'e Paris Diderot, 61, Avenue de l’Observatoire 75014, Paris France\\
$^7$ IAA-CSIC, Glorieta de la Astronom\'ia s/n, 18008 Granada, Spain\\
$^8$ Centro de Estudios de F\'isica del Cosmos de Arag\'on, Plaza San Juan 1, 44001 Teruel, Spain\\
$^9$ Instituto de Astrof\'isica de Canarias, V\'ia L\'actea s/n, 38200 La Laguna, Tenerife, Spain\\
$^{10}$ Departamento de Astrof\'isica, Facultad de F\'isica, Universidad de La Laguna, 38206 La Laguna, Spain\\
$^{11}$ Observat\'orio Nacional-MCT, Rua Jos\'e Cristino, 77. CEP 20921-400, Rio de Janeiro-RJ, Brazil\\
$^{12}$ Department of Theoretical Physics, University of the Basque Country UPV/EHU, 48080 Bilbao, Spain\\
$^{13}$ IKERBASQUE, Basque Foundation for Science, Bilbao, Spain\\
$^{14}$ Departamento de F\'isica At\'omica, Molecular y Nuclear, Facultad de F\'isica, Universidad de Sevilla, 41012 Sevilla, Spain\\
$^{15}$ Institut de Ci\`encies de l'Espai (IEEC-CSIC), Facultat de Ci\`encies, Campus UAB, 08193 Bellaterra, Spain\\
$^{16}$ Departamento de Astronom\'ia, Pontificia Universidad Cat\'olica. 782-0436 Santiago, Chile\\
$^{17}$ Instituto de Astronom{\'{\i}}a, Geof{\'{\i}}sica e Ci\'encias Atmosf\'ericas, Universidade de S{\~{a}}o Paulo, S{\~{a}}o Paulo, Brazil\\
$^{18}$ Departamento de Matem\'atica Aplicada y Estad{\'{\i}}stica, Universidad Polit\'ecnica de Cartagena, C/Dr. Fleming s/n, 30203 Cartagena, Spain\\
$^{19}$ Instituto de F\'{\i}sica Te\'orica, (UAM/CSIC), Universidad Aut\'onoma de Madrid, Cantoblanco, E-28049 Madrid, Spain\\
$^{20}$ Campus of International Excellence UAM+CSIC, Cantoblanco, E-28049 Madrid, Spain\\
$^{21}$ APC, AstroParticule et Cosmologie, Universit\'{e} Paris Diderot, CNRS, 10, rue Alice Domon et L\'{e}onie Duquet, 75013 Paris, France\\
}

\end{document}